%% file: paper.tex
\documentclass[11pt]{article}
\usepackage[margin=0.5in]{geometry}
\usepackage{todonotes}
\usepackage{graphicx}
\usepackage{listings}
\usepackage{color}
\usepackage{hyperref}
\usepackage{multirow}
\usepackage{pifont}
\usepackage{xcolor,colortbl}
\usepackage{hyperref}
\hypersetup{colorlinks=true,linkcolor=blue!50!black,citecolor=blue!75!black,filecolor=black,urlcolor=blue!75!black}
\usepackage{amsmath}
\usepackage{amssymb}
\usepackage{xspace}
\usepackage{makecell}
\usepackage{ulem} 
\usepackage[frozencache=true,cachedir=minted-cache]{minted} 
\usepackage[htt]{hyphenat}      
\usepackage{subcaption}
\usepackage{graphicx,import}
\usepackage{booktabs}

\newcommand{\cross}[0]{\cellcolor{red!65}\ding{53}}
\newcommand{\valid}[0]{\cellcolor{green!75!black}\ding{51}}
\newcommand{\warn}[0]{\cellcolor{orange!75}?}
\newcommand{\na}[0]{\cellcolor{gray!25}}
\newcommand{\s}[1]{\cellcolor{cyan!25}#1}

\newcommand{\pytracer}[0]{PyTracer\xspace}

\include{href}

\usepackage{tabularx} 
\makeatletter
\newcommand{\thickhline}{%
    \noalign {\ifnum 0=`}\fi \hrule height 1pt
    \futurelet \reserved@a \@xhline
}
\newcolumntype{"}{@{\hskip\tabcolsep\vrule width 1pt\hskip\tabcolsep}}
\makeatother

\lstdefinestyle{customPython}{
  belowcaptionskip=1\baselineskip,
  breaklines=true,
  xleftmargin=\parindent,
  language=Python,
  showstringspaces=false,
  basicstyle=\scriptsize\ttfamily,
  keywordstyle=\bfseries\color[rgb]{0.580, 0.000, 0.827},
  commentstyle=\itshape\color{green!40!black},
  identifierstyle=\bfseries\color{cyan!75!black},
  stringstyle=\color{orange},
  deletekeywords={double,float},
  classoffset=1, 
  otherkeywords={double,float},
  morekeywords={double,float},
  keywordstyle=\bfseries\color{green!55!black},
  classoffset=0
}

\lstdefinestyle{customC}{
  belowcaptionskip=1\baselineskip,
  breaklines=true,
  xleftmargin=\parindent,
  language=C,
  showstringspaces=false,
  basicstyle=\scriptsize\ttfamily,
  keywordstyle=\bfseries\color[rgb]{0.580, 0.000, 0.827},
  commentstyle=\itshape\color{green!40!black},
  identifierstyle=\bfseries\color{cyan!75!black},
  stringstyle=\color{orange},
  deletekeywords={double,float},
  classoffset=1, 
  otherkeywords={double,float},
  morekeywords={double,float},
  keywordstyle=\bfseries\color{green!55!black},
  classoffset=0
}

\begin{document}

\makeatletter
\let\orig@lstnumber=\thelstnumber
\newcommand\lstsetnumber[1]{\gdef\thelstnumber{#1}}
\newcommand\lstresetnumber{\global\let\thelstnumber=\orig@lstnumber}
\makeatother

\title{PyTracer: Automatically profiling numerical instabilities in Python}
\author{Yohan Chatelain$^1$, Nigel Yong$^1$,  Gregory Kiar$^2$,  Tristan Glatard$^1$\\
$^1$Department of Computer Science and Software Engineering, Concordia University, Montreal, Canada\\
$^2$Center for the Developing Brain, Child Mind Institute, New York, NY, USA}
\date{\textit{This work has been submitted to the IEEE for possible publication. 
Copyright may be transferred without notice, after which this version may no 
longer be accessible.}}
\maketitle

\begin{abstract}
Numerical stability is a crucial requirement of reliable scientific computing. However, despite the pervasiveness of Python in data science, analyzing large Python programs remains challenging due to the lack of scalable numerical analysis tools available for this language. To fill this gap, we developed \pytracer, a profiler to quantify numerical instability in Python applications. \pytracer transparently instruments Python code to produce numerical traces and visualize them interactively in a Plotly dashboard. We designed \pytracer to be agnostic to numerical noise model , allowing for tool evaluation through Monte-Carlo Arithmetic, random rounding, random data perturbation, or structured noise for a particular application. We illustrate \pytracer's capabilities by testing the numerical stability of key functions in both SciPy and Scikit-learn, two dominant Python libraries for mathematical modeling. Through these evaluations, we demonstrate \pytracer as a scalable, automatic, and generic framework for numerical profiling in Python.
\end{abstract}

\section{Introduction}

The scientific Python ecosystem has become a central component of data
analyses in recent years, owing to its rich offering of data manipulation, array programming,
numerical analysis, and visualization tools. In particular, libraries such
as NumPy~\cite{harris2020array}, SciPy~\cite{virtanen2020scipy}, scikit-learn~\cite{pedregosa2011scikit} or PyTorch~\cite{paszke2019pytorch} are used in hundreds of publications each year, and serve as a reference set of high-quality open-source core scientific libraries. Researchers have built numerous domain-specific software tools from this ecosystem, leveraging Python's simplicity and flexibility. 

Numerical stability is a crucial requirement of reliable scientific data
processing. In unstable analyses, small numerical perturbations introduced by data noise, software and hardware updates, or parallelization have potential to introduce substantial deviations in final results, leading to potentially different scientific conclusions, ultimately threatening the reliability of computational analyses. To prevent this, numerical stability analyses need to be conducted systematically, however, no practical tool currently exists to conduct such analyses for Python programs.

We present PyTracer, a numerical stability profiling and visualization tool for Python programs.
\pytracer adopts a dynamic approach that evaluates numerical stability using program execution traces, and is applicable to large programs without manual intervention. The transparent nature of \pytracer is particularly valuable in contrast to theoretical approaches based on condition numbers or backward error analysis, which require detailed modeling of the program and its numerical implementation. Similarly, static code analysis techniques hardly scale to large code bases, particularly in dynamically-typed languages such as Python.

\pytracer estimates the variability of floating-point variables by combining program traces obtained with different numerical perturbations, which requires (1) generating numerically-perturbed executions, (2) tracing floating-point computations along program executions, and (3) visualizing stability evaluations. \pytracer addresses these requirements with (1) a ``fuzzy" Python interpreter instrumented with Monte-Carlo arithmetic, (2) a dynamic instrumentation of Python functions, modules, and classes through meta-programming, (3) an interactive Plotly dashboard to visualize stability measures throughout program executions.

This manuscript presents the design of \pytracer and demonstrates its potential in various use cases. 
Through stability evaluations of the
SciPy and scikit-learn libraries, we demonstrate \pytracer as a practical solution to review the numerical stability of extensive code bases. 

\section{\pytracer design}

\begin{figure}
    \centering
    \includegraphics[width=\linewidth]{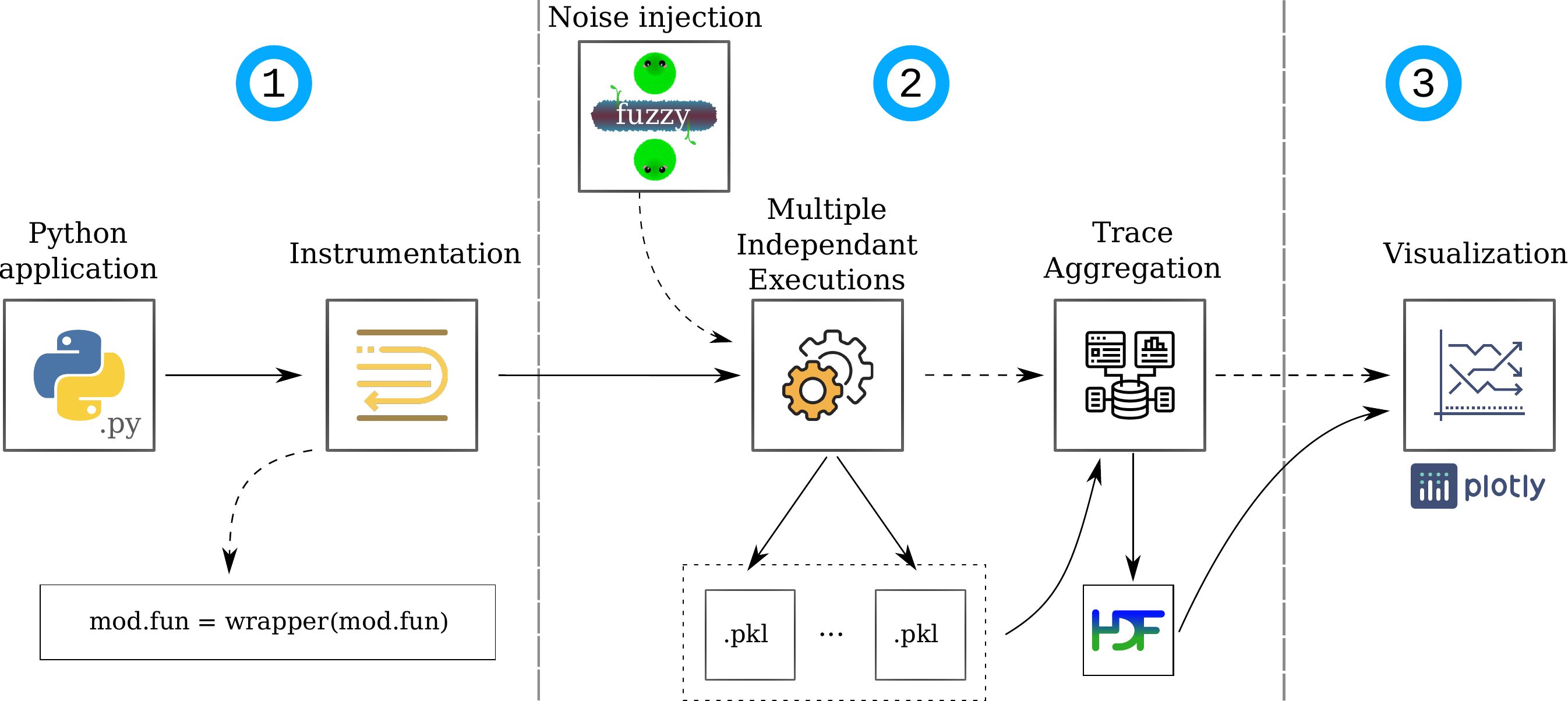}
    \caption{Pytracer workflow
    }
    \label{fig:workflow}
\end{figure}

\pytracer adopts the three-step workflow shown in Figure~\ref{fig:workflow}. First, \pytracer uses \textit{monkey patching} to replace the functions called by the application with a wrapper that saves their arguments and returned values in a trace file.
Second, it executes the application multiple times with noise injection, and aggregates the resulting trace files to compute summary statistics about each traced variable. Finally, a Plotly dashboard provides interactive visualizations highlighting unstable code sections.


\subsection{Python code instrumentation}


\pytracer dynamically instruments Python modules
without requiring source code modifications in the application.
To do so, it creates a new, instrumented version of each module that preserves its original attributes so that the application can transparently use it.
By default, \pytracer instruments all the attributes
in a given module except the special ones of the form \texttt{\_\_*\_\_}, which is common for reserved Python methods.
However, some attributes are not writable. In this case, \pytracer preservers the original attributes and warns the user. The user can also restrict the list of traced attributes through
an inclusion/exclusion mechanism. The main advantage of this instrumentation approach is the lack of required modification in the application source code, in contrast with decorator-based approaches. In addition, this technique is scalable and applicable in read-only environments such as Singularity containers or HPC clusters.

\subsubsection{Intercepting module import}

Python loads modules through an import mechanism triggered by the \texttt{import} keyword that involves two objects: \texttt{finders} and \texttt{loaders}. \texttt{Finders} search for packages (or 
namespaces\footnote{Python distinguishes between \texttt{regular} and \texttt{namespace} packages.
A regular package is a directory that contains a \_\_init\_\_.py file and potentially subdirectories (sub-packages) 
that contain themselves a \texttt{\_\_init\_\_.py} file and so on recursively. 
The package hierarchy follows those of the directory. 
Namespace packages introduced in Python 3.3 (\href{https://www.python.org/dev/peps/pep-0420/}{PEP 420}) do not contain an
\_\_init\_\_.py file and allow for flexible directory structures. Hence, parts of the package can be located in zip files, on the network, or in separated directories. There are no functional differences between both types of packages.}) on the local storage --- such as standard packages installed through the pip package manager --- or on a distant server.
\texttt{Finders} do not load modules. Instead, they return a specification (\texttt{spec} object) that includes 
information on where to find and how to load modules.
\texttt{Loaders} create module objects, initialize them, and fill import-related attributes 
such as \texttt{\_\_spec\_\_}. 
\texttt{Loaders} then execute modules and populate their namespace. Finally, modules are bound to their import name in the \texttt{sys.modules} dictionary.

Python supports custom \texttt{finder} classes registered in the \texttt{sys.meta\_path} list.
When Python encounters an \texttt{import} statement, it first looks for an existing binding in \texttt{sys.modules} and then iterates over the \texttt{finder} classes in the \texttt{sys.meta\_path} list until it finds the module. \pytracer adds a custom \texttt{finder} class at the head of \texttt{sys.meta\_path} that intercepts
the module import and creates the module with a custom \texttt{loader} class.

\pytracer's \texttt{loader} class first loads the original module, then copies it as a new instance of the \texttt{ModuleType} class. It then calls the appropriate instrumentation function for each attribute depending on its type (function, class, or instances) and replaces the original module in the \texttt{sys.modules} map with the instrumented version. Finally, it updates all existing references to the original module contained in the global symbol table.
As a result, the application will transparently call the instrumented module.



\subsubsection{Instrumenting module functions}

Listings~\ref{fig:wrapper_creation} and~\ref{fig:generic_wrapper} show \pytracer's function instrumentation mechanism.
For each function to instrument, \pytracer's \texttt{loader} class creates a wrapper function, dynamically compiles it with the \texttt{compile} builtin function, and substitutes it for the original module function.  Although simple, this technique does not apply to callable class instances, i.e., class instances that have a \texttt{\_\_call\_\_} attribute and are not of type \texttt{FunctionType}. Indeed, applying the wrapper technique would 1) modify the type of the class instance to \texttt{FunctionType}, which could cause syntax and semantic errors, and 2) mask class attributes, leading to \texttt{AttributeError} exceptions.
\pytracer overrides the \texttt{\_\_call\_\_} attribute with the wrapper function when possible to overcome this issue.
When the \texttt{\_\_call\_\_} attribute is not writable,  Pytracer does not instrument the class and returns a warning.

\begin{listing}
    \centering
\begin{lstlisting}[language=Python,style=customPython]
def get_wrapper_function(module, qualname, name, function):
    """Returns the instrumented function as a string.
    This string will be compiled with the compile builtin function
    
    Parameters:
        module: Name of the function module
        qualname: Qualified name of the function
        name: Name of the function
        function: Python object function
        
    Returns:
        wrapper_code: The code source of the wrapper
    """
    function_id = id(function)
    info = (function_id, function_module, function_qualified_name)
    wrapper_code = f"""
    def {function_wrapper_name}(*args, **kwargs):
        return generic_wrapper({info},*args,**kwargs)"""
    return wrapper_code
\end{lstlisting}
    \caption{Function to create the instrumented version of a function.
    \pytracer uses the identifier of the function instead of the 
    actual function to handle aliases and avoid duplicate instrumentation.}
    \label{fig:wrapper_creation}
\end{listing}

\begin{listing}
    \centering
\begin{lstlisting}[language=Python,style=customPython,]
def generic_wrapper(self, info, *args, **kwargs):
    """Generic wrapper saving inputs and outputs of the wrapped function.
    The id_dict dict keeps a mapping between the original function identifier
    and itself. Arguments are binded to avoid mispositioning.
    Returns the actual output of the wrapped function.
    
    Parameters:
        info: Tuple with the function's id, function's module and function's name
        *args: Positional calling arguments
        **kwargs: Keyword calling arguments
    
    Returns:
        outputs: Outputs of the wrapped function
    """
    fid, fmodule, fname = info
    function = original_function_cache.id_dict[fid]
    bind = Binding(function, *args, **kwargs)
    stack = self.backtrace()
    time = elements()
    self.inputs(time=time,
                module_name=fmodule,
                function_name=fname,
                function=function,
                args=bind.arguments,
                backtrace=stack)
    outputs = function(*bind.args, **bind.kwargs)
    self.outputs(time=time,
                 module_name=fmodule,
                 function_name=fname,
                 function=function,
                 args=outputs,
                 backtrace=stack)
    return outputs
\end{lstlisting}
    \caption{\pytracer's generic wrapper function that saves
    the inputs and outputs of the wrapped function.}
    \label{fig:generic_wrapper}
\end{listing}

\subsubsection{Instrumenting classes}

\pytracer instruments classes by wrapping their callable attributes as described previously. 
This instrumentation preserves class types, which avoids type mismatches in the instrumented application.
However, classes that describe particular base types and NumPy's universal functions (class \texttt{ufunc}) contain read-only attributes that cannot be instrumented.
By default, \pytracer returns a warning when it encounters such classes. 

\subsubsection{Instrumenting class instances}

\pytracer automatically instruments class instances since it instruments classes. However, it cannot instrument or instantiate classes that have read-only attributes --- in particular class \texttt{ufunc}, --- which is problematic given their pervasiveness in scientific computing. In particular, NumPy's universal functions include
vectorized elementary mathematical and other prevalent functions. 
Universal functions are implemented in C, which prevents their modification.

To overcome this issue, \pytracer wraps the class instance into a transparent wrap class (\texttt{twc}) that overloads
1) function \texttt{\_\_getattribute\_\_}, called when an instance attribute is accessed --- instrumented to return the attributes from the original class, and 2) function \texttt{\_\_call\_\_}, used when an instance behaves like a function through the () operator --- instrumented with function \texttt{generic\_wrapper}.
The \texttt{twc} returns the queried attribute of the instance instead of returning its attributes, allowing transparent access to the instance.  Finally, \pytracer overloads the builtin \texttt{type} function to preserve the type of the wrapped instance.

\subsubsection{Instrumenting iterators}

In functional programming, iterators traverse containers lazily, meaning that the next element in the sequence is only computed when the application uses it. This technique allows for the manipulation of virtually infinite sequences with finite memory. However, it implies that no complete mapping of the container returned by an iterator exists in memory since the application computes each element on the fly. Therefore, iterators are not serializable which implies that \pytracer cannot save them to output traces. 
A workaround would be to convert iterators into explicit containers, however, this would increase the memory footprint significantly. \pytracer, therefore, does not instrument non-builtin iterators.


\subsection{Detecting numerical instabilities}

\pytracer detects numerical instabilities by computing summary statistics across multiple executions perturbed with numerical noise. While \pytracer's instrumentation, trace aggregation, and visualization work with various types of numerical noise, we experimented primarily with Monte-Carlo Arithmetic (MCA) as it is an accurate model for floating-point errors.

\subsubsection{Noise injection}
\label{sec:fuzzy}

Three  main  approaches  exist  to  detect  numerical  instabilities:  stochastic  arithmetic,  uncertainty  or  sensitivity analysis [23], and random seed analysis [11].  PyTracer is agnostic to the noise model used and therefore works for all methods mentioned.  For our use cases, we focus on stochastic arithmetic since it does not make any assumption about the noise model, in contrast to sensitivity analysis

Stochastic arithmetic leverages randomness to estimate numerical instabilities coming from the use of floating-point representations. The main idea is to treat round-off errors as random variables and characterize them statistically.
Monte Carlo Arithmetic (MCA)~\cite{parker1997monte} uses this principle by introducing two improvements:
(i) a virtual precision parameter, allowing to simulate reduced working precision, and (ii) different perturbation modes to introduce perturbations
on function inputs or output.
MCA replaces each floating-point number $x$ with a stochastic counterpart computed as:
\[
inexact(x) =  x + \beta^{e_x - t}\xi
\]
where $\beta$ is the number base, $e_x$ is the magnitude of $x$, $t$ the virtual precision and $\xi \in (-\frac{1}{2},\frac{1}{2})$ is a random uniform variable.
Virtual precision allows for the simulation of reduced working precision.
MCA can be applied in three modes: Random Rounding (RR), Precision Bounding (PB), and full MCA, which apply function $inexact$ to the output, the inputs, or both, respectively. While the RR mode is equivalent to stochastic rounding, the PB mode can also identify catastrophic cancellations.

Stochastic arithmetic quantifies numerical error using the number of significant digits $s$ estimated among the sampled values. A common formula to determine this number from MCA samples is presented in~\cite{parker1997monte}:
\begin{equation}
s = -\log_{\beta}{ \left| \dfrac{\sigma}{\mu} \right|} \label{eq:sig-digits}
\end{equation}
where $\mu$ and $\sigma$ are the sample mean and standard deviation of a variable sampled with MCA.  
Sohier et al.~\cite{sohier2018confidence} recently provided a generalization of this formula to include confidence intervals.

We enable MCA in Python programs via Verificarlo~\cite{verificarlo}, a clang-based compiler~\cite{lattner2008llvm} that replaces floating-point operations by a generic call to a configurable floating-point model. Several floating-point models are available~\cite{chatelain2019automatic,chatelain2019outils}, also called backends.
\pytracer leverages ``fuzzy"~\cite{kiar2020comparing}, a collection of software tools compiled with Verificarlo. In particular, fuzzy provides MCA-instrumented versions of the Python interpreter as well as the BLAS, LAPACK, NumPy, SciPy, and scikit-learn libraries, which enables MCA for a wide range of existing scientific Python programs. Fuzzy is available in Verificarlo's GitHub organization at \href{https://github.com/verificarlo/fuzzy}{\url{github.com/verificarlo/fuzzy}}.


\subsubsection{Trace format}

\pytracer stores traces in the pickle format, a binary format to serialize Python objects.
The main advantages of the pickle format compared to other serialization approaches such as marshal or JSON are its ability to serialize most Python objects and its native compression.  In addition, pickle writes Python objects sequentially, which preserves the temporal ordering required by subsequent trace analyses.

Traces store numerical values at the granularity of function calls. When the application invokes a function, \pytracer saves a call input object containing contextual information (function's name, module's function, stacktrace) and the values of the function arguments. When the function returns, \pytracer saves a similar call output object containing the returned value.

\subsubsection{Trace aggregation}

Once traces are available, \pytracer sequentially parses them and computes the mean, standard deviation, and the number of significant digits (as in Equation~\ref{eq:sig-digits}) of all saved values. \pytracer saves these summary statistics in an HDF5 file, a hierarchical format that facilitates visual browsing since only
required information are loaded when requested by the user.
This operation critically relies on the ordering of call input and output objects in the traces. Therefore, it assumes that the application is single-threaded and that the control flow does not depend on a random state. 

\begin{figure}
    \centering
    \includegraphics[width=\textwidth]{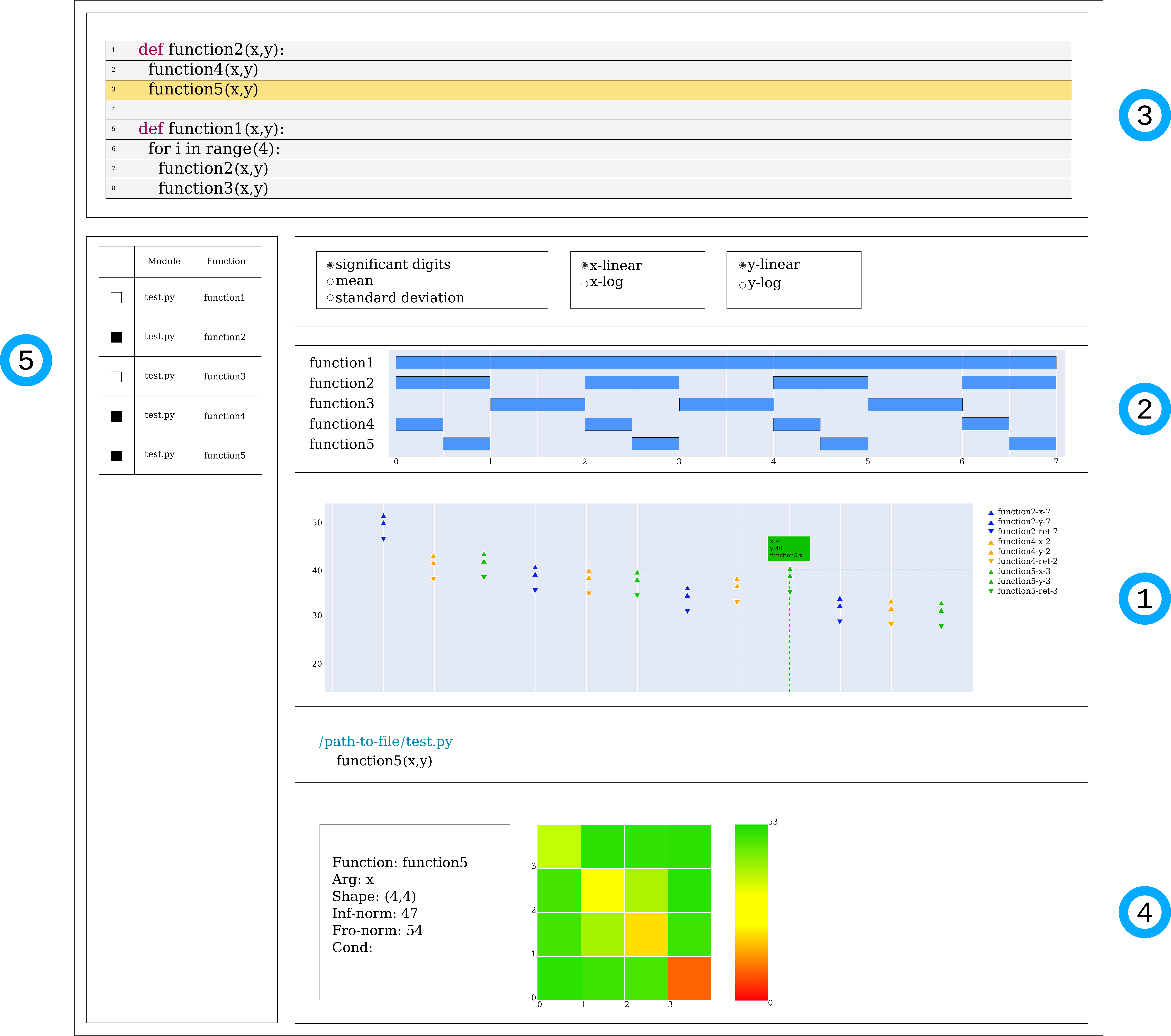}
    \caption{Pytracer visualization layout.}
    \label{fig:visu-layout}
\end{figure}

To aggregate traces, \pytracer successively unpickles call objects in each trace, checks
that the call objects refer to the same function and stacktrace, and terminates the parsing otherwise.
PyTracer stores function arguments or returned values in a NumPy array for each matching call object to compute statistics.
For compound objects such as tuples, dictionaries, or objects, it recursively parses the fields until finding
numeric types or NumPy arrays of numeric types.

\subsection{Interactive visualization}
The third main component of \pytracer is its interactive trace visualizer built with Plotly Dash~\cite{plotly}, a Python framework for web visualization.
Plotly Dash abstracts the low-level details of web applications in a highly customizable development interface allowing to offload heavy computations to a server if necessary.
Figure~\ref{fig:visu-layout} presents the general layout of the \pytracer visualizer, divided into five sub-components:
\begin{enumerate}
 \item Timeline view: This view displays the values of each input and output for a given function at a given invocation. By default, \pytracer shows the number of significant digits computed across the traces. The mean and standard deviation across perturbed samples are also available. 
 The visualizer uses an upper triangle for inputs and a lower triangle for outputs. \pytracer internally represents all values as NumPy arrays. In the case of non-scalar values or types that cannot be trivially cast into NumPy arrays, \pytracer splits the value into several variables. Hence, \pytracer will transform a class containing $n$ floating-point parameters into $n$ variables. Each variable is prefixed with the original value and postfixed with the name of the parameter.
    Finally, non-scalar values (vectors or matrices) are represented by their average and can be inspected with the Zoom view (see below).
 \item Gantt chart view: This view displays the call graph as a Gantt chart representing the instrumented functions. Each function call is represented with a bar.
    The time overlaps reflect the caller/callee relation.
    For example, in the figure, \texttt{function1} calls \texttt{function2} that itself calls \texttt{function4} and \texttt{function5}. 
    This view provides insights on the calling context for each function, which may influence numerical stability.
 \item  Source code view: This view displays a local snapshot of the source code file of the variable hovered in the timeline view. 
 \item Zoom view: This view provide details about non-scalar values. For example, a matrix may have instabilities around its diagonal but not in other regions.
    The visualizer also provides additional metrics such as array shape, infinite norm, or condition number.
\item  Functions list view: This view allows the user to select the functions plotted in the timeline view.
\end{enumerate}

\section{Numerical evaluations}

We evaluated \pytracer on two widely-used Python libraries: SciPy~\cite{virtanen2020scipy}, the main library for scientific computing in Python, and  scikit-learn~\cite{pedregosa2011scikit}, a reference machine-learning framework that
 implements a wide range of state-of-the-art algorithms. For each library, we executed the software examples provided in their respective GitHub repositories. We traced each example five times, having activated MCA in the Python interpreter, BLAS, LAPACK, NumPy, SciPy, and scikit-learn. We repeated the experiment with two MCA modes, Random Rounding (RR) and Full MCA, resulting in 2 experiments for each example. We used the default virtual precision in Verificarlo, namely 24 bits for single-precision floating-point values and 53 bits for double-precision values, which simulates machine error. We used Python-3.8.5, NumPy-1.19.1, SciPy-1.5.4, and scikit-learn-0.23.2.

\subsection{Results overview}

Table~\ref{tab:pytracer_results_summary} summarizes
\pytracer's outcome on SciPy and scikit-learn examples.
Columns \textit{trace} and \textit{parse} report on the tracing and aggregation steps with three possible outcomes: 1) green for success, meaning that the execution ended well and no issues were raised by the application, 2) red for numerical error in the application, meaning that the execution raised a Python error before the end, and 3) orange for errors due to the fuzzy noise injection that are not considered numerical errors, meaning that the execution raised a Python error but due to a perturbation that should not have been introduced (i.e. a floating-point value representing an integer).
Overall, the tracing and parsing of SciPy and scikit-learn examples in Random Rounding mode was entirely successful for 32/40 examples and showed an average precision loss of 8 bits. The remaining examples failed due to numerical errors in the libraries, which we discuss hereafter. The instrumentation with Full MCA is more invasive and reduced the number of successful executions to 12/40. Among the 28/40 failed executions, one failed due to an actual numerical error (\texttt{Bayesian Ridge Regression} example) and the other ones failed due to 
perturbations that should not have been introduced.

Column \textit{sigbits} in Table~\ref{tab:pytracer_results_summary} summarizes the precision for the main outputs of SciPy and scikit-learn examples. 
We measured the precision in significant bits on the final primary example output by taking the element-wise mean for non-scalar values and, if the example produces many outputs, the average value among all of them. The numerical quality across all examples varies from 0 to 52, with an average of 44 significant bits. For both modes, we note that only two examples failed due to a numerical error 
(\texttt{face\_recognition} for RR and \texttt{Bayesian Ridge Regression} for Full MCA).

\begin{table}[]
    \centering
    \small
    \begin{subfigure}[t]{\linewidth}
    \centering
    \begin{tabular}{|lll|c|c|c|c|c|c|}
    \hline
    \multicolumn{3}{|c}{ \multirow{2}{*}{Application} } & \multicolumn{3}{|c|}{Random Rounding (RR)} & \multicolumn{3}{c|}{Full MCA} \\
    \cline{4-9}
    & & & trace & parse & sigbits & trace & parse & sigbits \\
    \hline
    
    \multicolumn{1}{|c|}{ \multirow{20}{2em}{ \rotatebox{90}{SciPy} } } &   
    \multicolumn{1}{c|}{ \multirow{3}{2em}{ \rotatebox{90}{FFT} }}  & \discreteCosineRF &  \valid & \valid & \s{52} & \warn & \cross & \s{47} \\ 
    \multicolumn{1}{|c|}{} & \multicolumn{1}{c|}{} & \fftOneDRf & \valid & \valid & 41 & \warn  & \cross & - \\
    \multicolumn{1}{|c|}{} & \multicolumn{1}{c|}{} & \fftTwoDRf & \valid & \valid & \s{52} & \valid & \valid & \s{52} \\
    \cline{2-9}
    \multicolumn{1}{|c|}{} &  \multicolumn{1}{c|}{  \multirow{5}{*}{ \rotatebox{90}{Interpolation} } }
    & \interOneDRf & \valid & \valid & 51 & \warn & \cross &  - \\
    \multicolumn{1}{|c|}{} & \multicolumn{1}{c|}{} & \multiRf      & \valid & \valid & \s{51} & \warn  & \cross & \s{-} \\
    \multicolumn{1}{|c|}{} & \multicolumn{1}{c|}{}  & \bsplineRf    & \valid & \valid &10 &  \valid & \valid & 10\\
\multicolumn{1}{|c|}{} & \multicolumn{1}{c|}{} &     \splineOneDRf & \valid & \cross & \s{52} & \warn  & \cross & \s{52} \\
\multicolumn{1}{|c|}{} & \multicolumn{1}{c|}{} &     \splineTwoDRf & \valid & \valid & 10 &  \warn  & \cross & - \\
    \cline{2-9}
    \multicolumn{1}{|c|}{} & \multicolumn{1}{c|}{ \multirow{12}{2em}{ \rotatebox{90}{Optimization} } }
    & \bfgsRf & \valid & \valid & \s{48}  & \valid & \valid & \s{46} \\
\cline{3-9}
\multicolumn{1}{|c|}{} & \multicolumn{1}{c|}{} & \globalRf & \valid & \cross & 17 & \warn  & \cross & - \\
\multicolumn{1}{|c|}{} & \multicolumn{1}{c|}{} & | SHGO\footnote{Simplicial Homology Global Optimization} & \na & \na  & \s{25} & \na  & \na & \s{-} \\
\multicolumn{1}{|c|}{} & \multicolumn{1}{c|}{} & | Dual Annealing & \na & \na  & 4 & \na & \na  & - \\
\multicolumn{1}{|c|}{} & \multicolumn{1}{c|}{} & | Differential Evolution & \na & \na & \s{0} & \na & \na & \s{-} \\
\multicolumn{1}{|c|}{} & \multicolumn{1}{c|}{} & | Basin Hopping & \na  & \na & 40 & \na  & \na & - \\
\multicolumn{1}{|c|}{} & \multicolumn{1}{c|}{} & | SHGO Sobol    & \na  & \na  & \s{45} & \na  & \na & \s{-} \\
\cline{3-9}
\multicolumn{1}{|c|}{} & \multicolumn{1}{c|}{} & \slsqpRf  & \valid & \valid & 45 & \valid & \valid & 44 \\
\multicolumn{1}{|c|}{} & \multicolumn{1}{c|}{} & \lsmRf    & \valid & \valid &\s{48} & \warn  & \cross & \s{-} \\
\multicolumn{1}{|c|}{} & \multicolumn{1}{c|}{} & \nelderRf & \valid & \valid & 44 & \valid & \valid & 42 \\
\multicolumn{1}{|c|}{} & \multicolumn{1}{c|}{} & \ncgRf    & \valid & \valid & \s{48} & \valid & \cross & \s{47} \\
\multicolumn{1}{|c|}{} & \multicolumn{1}{c|}{} & \rootRf   & \valid & \cross & 52 & \warn  & \cross & 50 \\
\multicolumn{1}{|c|}{} & \multicolumn{1}{c|}{} & \rootLargRf      & \valid & \cross & \s{29} & \valid & \cross & \s{26} \\
\multicolumn{1}{|c|}{} & \multicolumn{1}{c|}{} & \rootLargePredRf & \valid & \cross & 42 & \valid & \cross & 42 \\
\multicolumn{1}{|c|}{} & \multicolumn{1}{c|}{} & \trustExactRf & \valid & \valid & \s{48}  & \valid & \valid & \s{46} \\
\multicolumn{1}{|c|}{} & \multicolumn{1}{c|}{} & \trustTruncRf & \valid & \valid & 48  & \valid & \valid & 46 \\
\multicolumn{1}{|c|}{} & \multicolumn{1}{c|}{} & \trustNCGRf   & \valid & \valid & \s{52}  & \valid & \valid & \s{47} \\
    \hline

    \multicolumn{2}{|c|}{ \multirow{20}{2em}{ \rotatebox{90}{Scikit-learn} } }
    & \AdaboostRf   & \valid & \valid & 48 & \warn  & \cross & - \\
    \cline{3-9}
    \multicolumn{2}{|c|}{} & \brrRf        & \valid & \valid & \s{25} & \cross & \cross & \s{-} \\
    \multicolumn{2}{|c|}{} & | $1^{st}$ dataset & \na  & \na & 25 & \na  & \na  & - \\
    \multicolumn{2}{|c|}{} & | $2^{nd}$ dataset & \na & \na  & \s{nan} & \na  & \na & \s{-} \\
    \cline{3-9}
    \multicolumn{2}{|c|}{} & \onlineClassifierComparisonRf & \valid & \valid & 20 & \warn & \cross & -  \\
    \multicolumn{2}{|c|}{} & \kmeansRf     & \valid & \valid & \s{52} & \valid & \cross & \s{-} \\
    \multicolumn{2}{|c|}{} & \covarianceRf & \valid & \cross & 48 & \warn  & \cross & - \\
    \multicolumn{2}{|c|}{} & \decisionRf   & \valid & \valid & \s{50} & \valid & \valid & \s{50} \\
    \multicolumn{2}{|c|}{} & \digitsRf     & \valid & \valid & 52 & \warn  & \cross & 52 \\
    \multicolumn{2}{|c|}{} & \faceRf       & \cross & \valid & \s{-}  & \warn  & \cross & \s{-} \\
    \multicolumn{2}{|c|}{} & \penaltyRf    & \valid & \valid & -  & \valid & \valid & - \\
    \multicolumn{2}{|c|}{} & \lassoRf      & \valid & \valid & \s{48} & \warn  & \cross & \s{47} \\
    \multicolumn{2}{|c|}{} & \hyperplaneRf & \valid & \valid & 3  & \valid & \cross & - \\
    \multicolumn{2}{|c|}{} & \mnistRf      & \valid & \valid & \s{43} & \warn  & \valid & \s{-} \\
    \multicolumn{2}{|c|}{} & \multitaskRf  & \valid & \valid & 50 & \warn  & \valid & 48 \\
    \multicolumn{2}{|c|}{} & \ompRf        & \valid & \valid & \s{52} & \valid & \valid & \s{52} \\
    \multicolumn{2}{|c|}{} & \pcaRf        & \valid & \valid & 48 & \valid & \valid & 48 \\
\cline{3-9}
    \multicolumn{2}{|c|}{} & \robustRf     & \valid & \valid & \s{50} & \warn  & \cross & \s{48} \\
    \multicolumn{2}{|c|}{} & | Linear        & \na & \na & 48 & \na  & \na & 46 \\
    \multicolumn{2}{|c|}{} & | RANSAC        & \na  & \na & \s{50} & \na  & \na  & \s{48} \\
\cline{3-9}
    \multicolumn{2}{|c|}{} & \toyRf        & \valid & \valid & 35 & \warn  & \cross & - \\
\cline{3-9}
    \multicolumn{2}{|c|}{} & \svmRf        & \valid & \valid & \s{43.5} & \valid & \cross & \s{-} \\
    \multicolumn{2}{|c|}{} & | Non-weighted  & \na & \na & 43 & \na & \na  & - \\
    \multicolumn{2}{|c|}{} & | Weighted      & \na  & \na & \s{44} & \na  & \na  & \s{-} \\
\cline{3-9}
    \multicolumn{2}{|c|}{} & \tomographyRf & \valid & \cross & - & \warn  & \cross & - \\
    \multicolumn{2}{|c|}{} & \weightedRf   & \valid & \valid & \s{44} & \valid & \cross & \s{-} \\
    \hline
    \end{tabular}
    \end{subfigure}
    \caption{\pytracer execution summary on SciPy and Scikit-learn examples. 
    \emph{trace}: outcome of \pytracer tracing, \emph{parse}: outcome of \pytracer trace parsing and aggregation, \emph{result}: precision for main outputs in significant bits. \valid: successful run, \warn: error from invalid noise injection, \cross: numerical error or divergent traces. \textit{nan} stands for (\texttt{Not-A-Number}). Example names are linked to their source code.}
    \label{tab:pytracer_results_summary}
\end{table}


\subsection{Effect of MCA modes}
\label{sec:impact_mca_modes}
Random Rounding perturbs only the result of a floating-point operation while Full MCA perturbs both the inputs and the result. 
Therefore, Full MCA may trigger catastrophic cancellations while Random Rounding only introduces round-off errors which generally have a lesser impact.
Moreover, Random Rounding preserves exact operations --- operations where the result can be exactly represented at the virtual precision --- while Full MCA does not.

\pytracer can only aggregate traces obtained from the same control flow path. In particular, program branches triggered by an unstable floating-point comparison lead to different control flows and result in parsing errors. 
For example, comparing a residual error or the distance between two points to a strict threshold in an iterative scheme generally results in unstable branches. 
Full MCA even perturbs exact values such as integers represented with floating-point values, leading to many execution failures. For instance, the \texttt{fft1} example from SciPy raised an \texttt{ValueError: operands could not be broadcast together with shapes (601,) (600,)} error due to an array shape mismatch. A closer inspection showed that one of the arrays involved in the error was created with the NumPy function \texttt{linspace} that itself calls function \texttt{arange} (Listing~\ref{fig:pyarray_range_code}). 
Lines 3174-3178 show that the function stores the array size in a floating-point value that will be perturbed by Full MCA, as shown in Table~\ref{tab:mca_result_linspace}, resulting in array sizes that fluctuate between N and N+1.
It is worth noting that 78\% of execution issues encountered in Full MCA mode instrumentation
are due to this type of ``typing" errors.  
Those code regions should not be perturbed by MCA since the injected noise does not make sense.

\begin{listing}
    \begin{lstlisting}[language=C,style=customC,numbers=left, firstnumber=3163, mathescape=true]
/*NUMPY_API
  Arange,
*/
NPY_NO_EXPORT PyObject *
PyArray_Arange(double start, double stop, double step /* default 1 */, int type_num) 
{
    npy_intp length; 
    PyArrayObject *range; $\lstsetnumber{\ldots}$
    ...$\lstresetnumber\setcounter{lstnumber}{3173}$
    double delta, tmp_len; $\lstsetnumber{\ldots}$
    ...$\lstresetnumber\setcounter{lstnumber}{3176}$
    delta = stop - start; 
    tmp_len = delta/step; $\lstsetnumber{\ldots}$
    ...$\lstresetnumber\setcounter{lstnumber}{3189}$
        length = _arange_safe_ceil_to_intp(tmp_len); $\lstsetnumber{\ldots}$
    ...$\lstresetnumber\setcounter{lstnumber}{3201}$
    range = (PyArrayObject *)PyArray_New(&PyArray_Type, 1, &length, type_num, $\lstsetnumber{}$
                        NULL, NULL, 0, 0, NULL);
    \end{lstlisting}
\caption{Source code of the \texttt{PyArray\_Range} Cython function called by NumPy function \texttt{linspace} to create an array of equally-spaced elements. The temporary array size assigned in line 3178 is stored as a floating-point value and is therefore perturbed in Full MCA mode, leading to differences amplified by the use of ceil rounding at line 3190 and resulting in different array sizes across MCA-perturbed executions. See also Table~\ref{tab:mca_result_linspace}.}
    \label{fig:pyarray_range_code}

\end{listing}

\begin{table}[]
    \centering
    \footnotesize
    \begin{tabularx}{{\textwidth}}{cccc>{\centering\arraybackslash}X>{\centering\arraybackslash}X>{\centering\arraybackslash}X}
                \thickhline
    \textbf{Mode}  & \textbf{start} & \textbf{stop} & \textbf{step} & \textbf{delta} & \textbf{tmp\_len} & \textbf{ceil(tmp\_len)}  \\
    \hline
    Exact & 0 & 600  & 1 & 600                             & 600                            & 600             \\
    RR    & 0 & 600  & 1 & 600 $\pm$ 0.0                   & 600 $\pm$ 0.0                  & 600 $\pm$ 0.0   \\
    MCA    & 0 & 600  & 1 & 599.9999999999999 $\pm$ 8e-14  & 600.0 $\pm$ 8e-14 &  600.01 $\pm$ 0.0995\\
            \thickhline

    \end{tabularx}
    \caption{Results obtained through lines 3177-3190 in Listing~\ref{fig:pyarray_range_code}. In contrast with RR, Full MCA does not preserve exact operations which leads to an array length oscillating between 600 and 601.}
    \label{tab:mca_result_linspace}
\end{table}

\subsection{SciPy}
\label{sec:scipy_tests}

We tested \pytracer on SciPy examples provided in the tutorial website section. 
SciPy is organized in several libraries targeting specific computing domains. We focused on \texttt{fftpack} (Fast Fourier Transform routines), \texttt{interpolate} (interpolation and smoothing splines), and \texttt{optimize} (Optimization and root-finding routines).

\subsubsection{FFT}

\begin{figure}
    \centering
    \includegraphics[width=\linewidth]{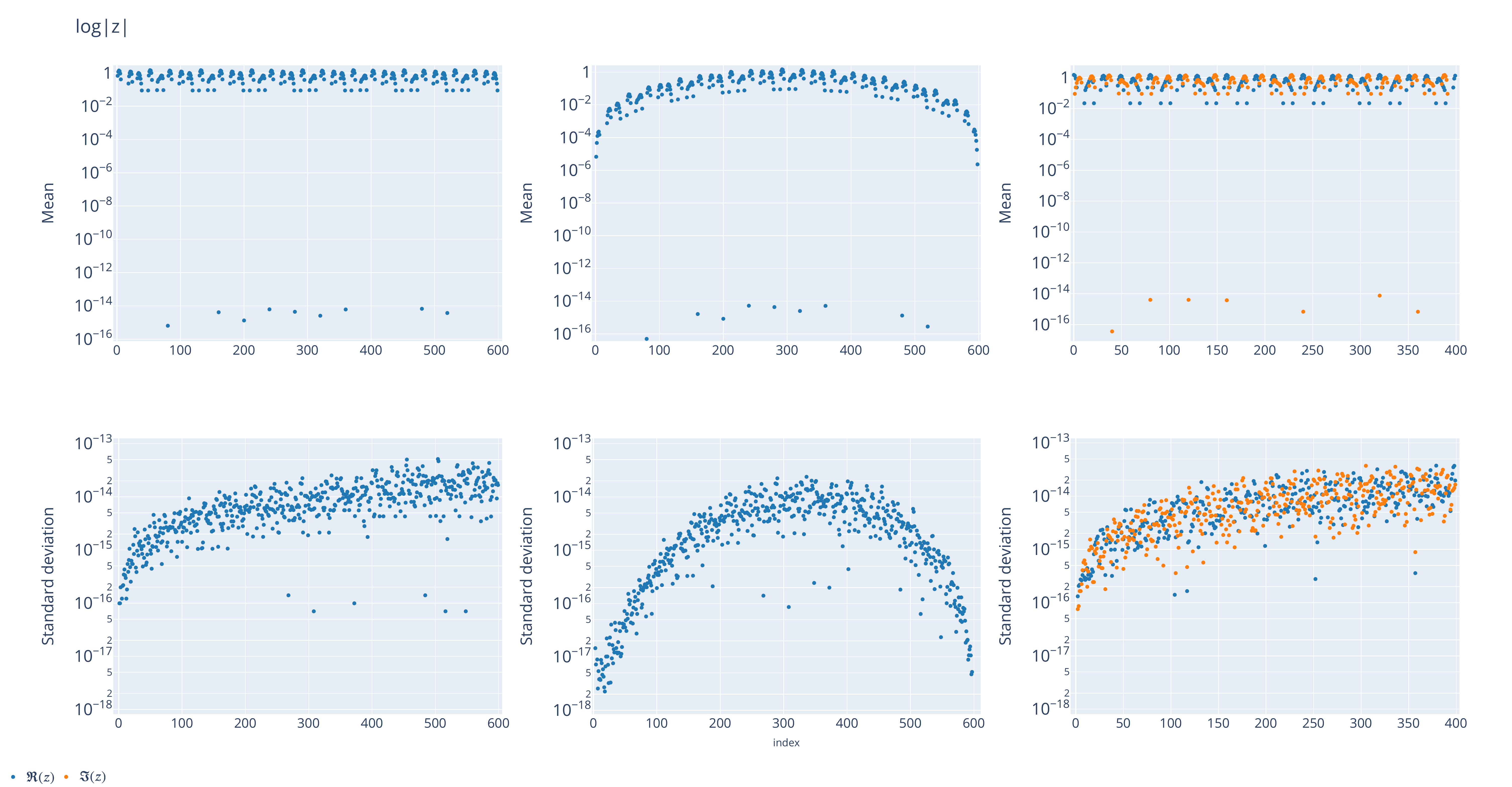}
    \caption{Absolute value of the mean and standard deviation values for 
    fft 1D inputs within RR mode (log scale). The real part is shown in blue and the imaginary part in orange.}
    \label{fig:fft1D_inputs}
\end{figure}

\begin{figure}
    \centering
    \includegraphics[width=\linewidth]{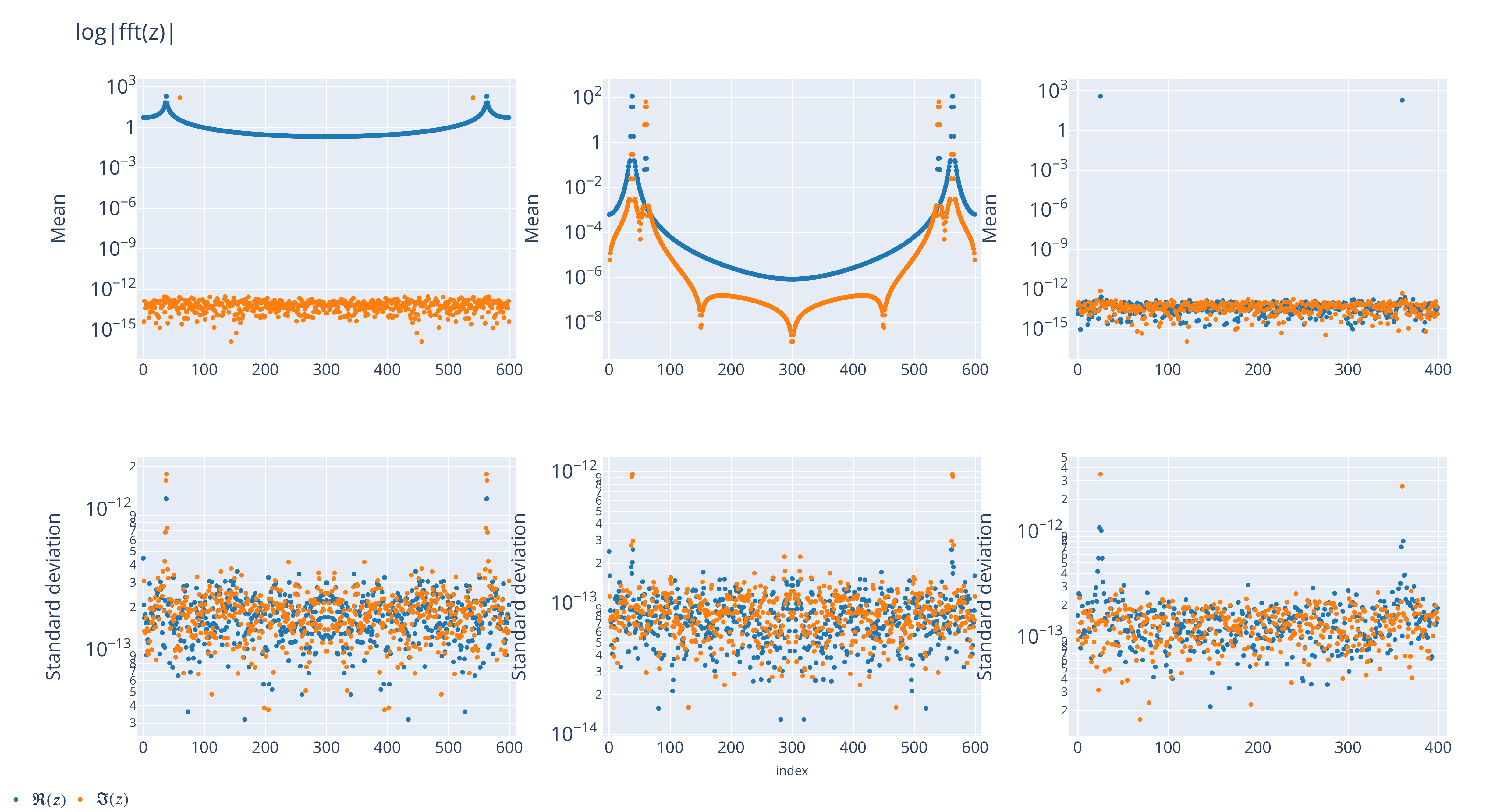}
    \caption{Absolute value of the mean and standard deviation values 
    for fft 1D outputs within RR mode (log scale).}
    \label{fig:fft1D_outputs}
\end{figure}

This package has three examples : \texttt{discrete\_cosine\_transform}, \texttt{fft\_1D}, and \texttt{fft\_2D}. All computations are done in double precision.
The results for \texttt{discrete\_cosine\_transform},\texttt{fft\_1D} and  \texttt{fft\_2D} show
precise numerical results with 48 significant bits on average. 
We observe that the FFT computation is stable to noisy inputs. Figures~\ref{fig:fft1D_inputs} (inputs) and~\ref{fig:fft1D_outputs} (outputs)
show the mean and standard deviation of the MCA samples. 
The three columns correspond to the FFT computations of: 
\begin{enumerate}
\item $z_1 = \sin(50 . 2\pi . x_i) + \dfrac{1}{2} \sin(80 . 2\pi . x_i),\; x_i = \frac{i}{600},\; i \in [0,600]$
\item $\mathrm{blackman}(400) \times z_1$
\item $ z_2= e^{50 . i 2\pi . x_i} + \dfrac{1}{2} e^{-80 . i2\pi .x_i },\; x_i = \frac{i}{800},\; i \in [0,400] $
\end{enumerate}

Top row in Figures~\ref{fig:fft1D_inputs} and~\ref{fig:fft1D_outputs} 
show the input mean value while the bottom row shows the standard deviation.
The points with low magnitude in Figure~\ref{fig:fft1D_inputs} correspond 
to inputs near 0 when the input of $\sin$ or 
$\exp$ is close to $k\pi$, $k \in \mathbb{Z}$.
Indeed, since MCA introduces a slight noise, the result is not exactly 0.
We can see in Figure~\ref{fig:fft1D_inputs} 
that the maximal noise introduced by MCA on the
inputs is in the order of $10^{-14}$, two orders of magnitude higher than the 
$ulp \simeq 10^-16$ for double precision. 
This slight noise introduced can be interpreted as white noise on the input signal. 
We can see on the bottom row of Figure~\ref{fig:fft1D_outputs} that the frequency noise is of 
the same order of of magnitude as the input noise, which is expected. 
The two peaks at x=38 and x=562 corresponds to the actual frequencies of the input signal.

\subsection{Interpolation}

This package has five examples: \texttt{interpolation\_1D}, \texttt{multivariate\_data}, \texttt{bspline}, \texttt{spline\_1D} and \texttt{spline\_2D}.

\texttt{interpolation\_1D} interpolates function $f(x)=\cos(\frac{-x^2}{9})$ with 11 points $x\in[0,10]$.
For the five interpolation methods tested (linear, cubic, nearest, previous, next), the solution found is precise, up to 51 bits out of the 53 available.

\texttt{multivariate\_data} interpolates the grid $f(x,y)=x(1-x)\cos(4\pi.x)  \sin(4\pi.y^2)^2$ with $(x,y) \in [0,1] \times [0,1]$. It samples 1,000 points for each coordinate and uses three interpolation methods: nearest, linear and cubic. Our results show a precision of 51 bits on average for the three methods. 

\begin{figure}
\centering
\begin{subfigure}{.3\linewidth}
    \includegraphics[width=\linewidth]{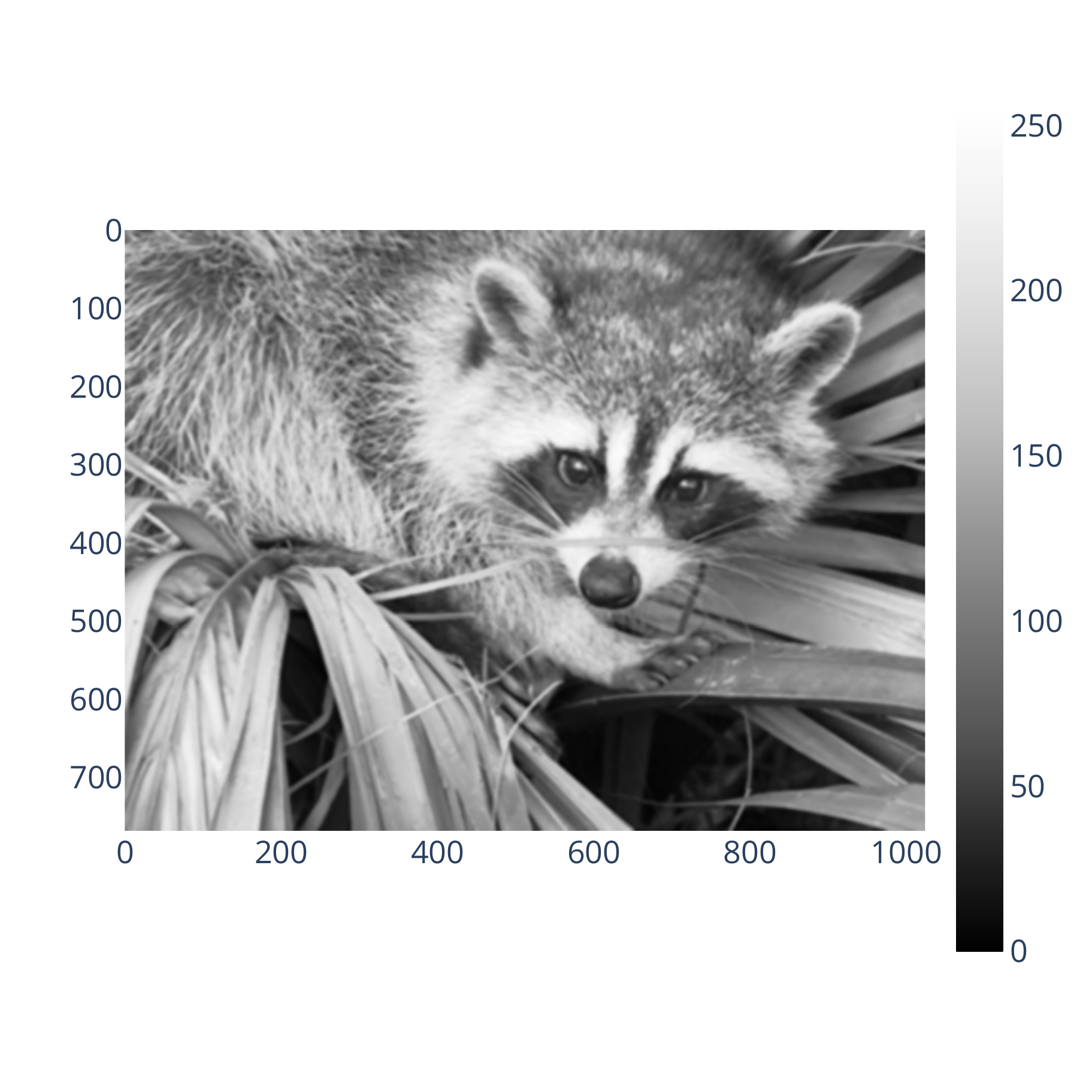}
    \caption{Original image.}
    \label{fig:bspline_original_image}
\end{subfigure}\\
\begin{subfigure}{0.3\linewidth}
    \includegraphics[width=\linewidth]{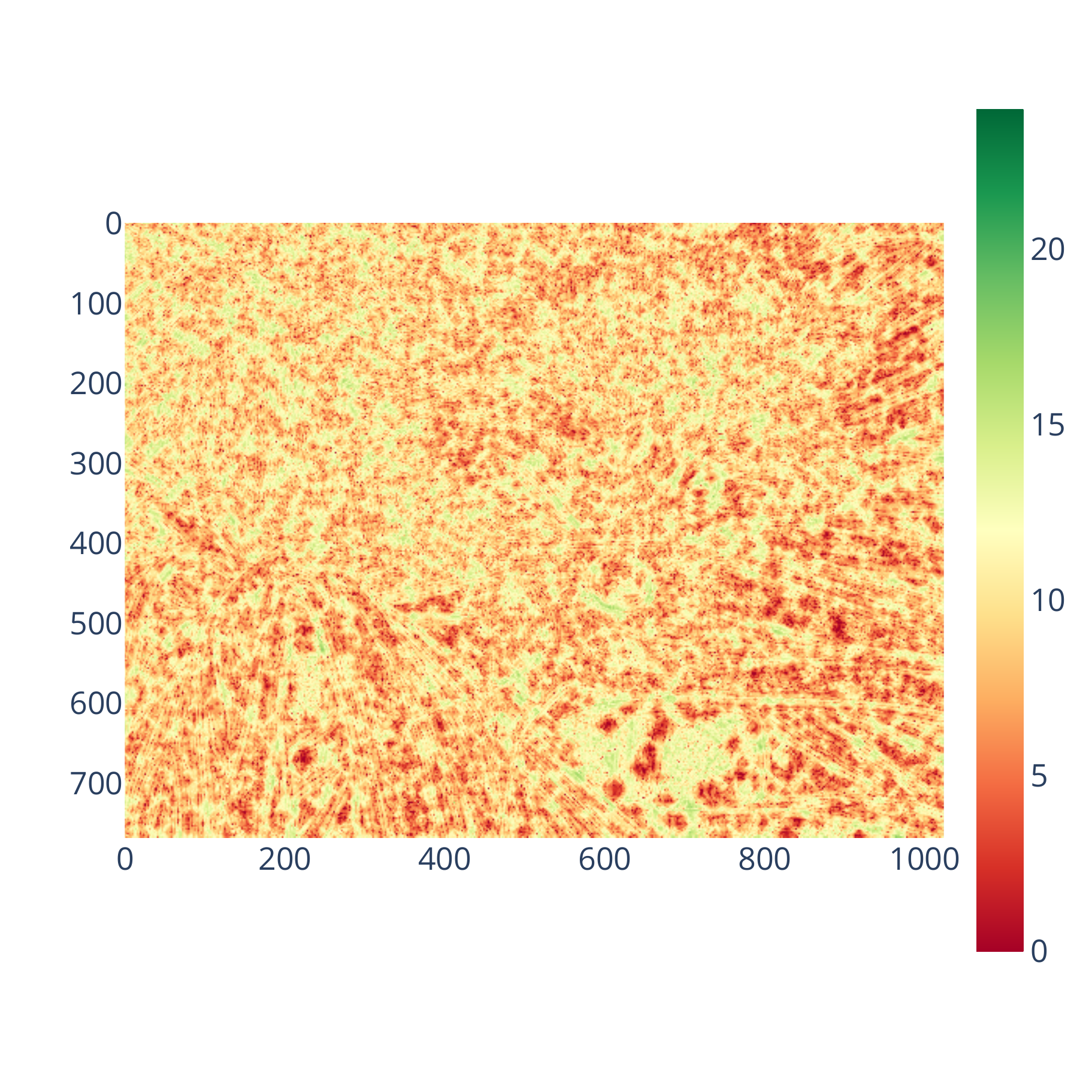}
    \caption{\centering\texttt{sepfir2d} significant bits. \newline \textcolor{white}{.}}
    \label{fig:bspline_bisplev_sig}
\end{subfigure}
\begin{subfigure}{0.3\linewidth}
    \includegraphics[width=\linewidth]{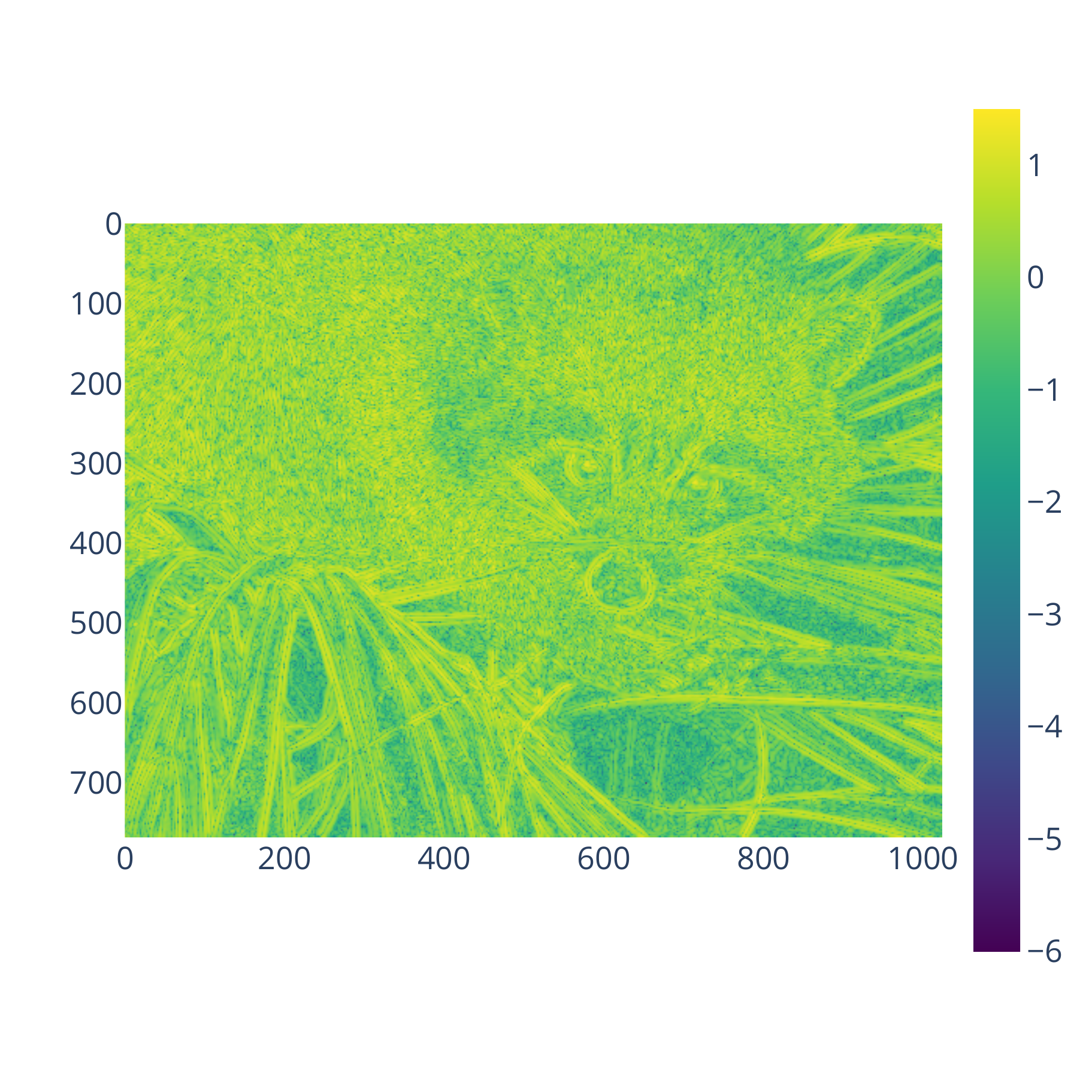}
    \caption{\centering\texttt{sepfir2d} mean. \newline (log)}
    \label{fig:bspline_bisplev_mean}
\end{subfigure}
\begin{subfigure}{0.3\linewidth}
    \includegraphics[width=\linewidth]{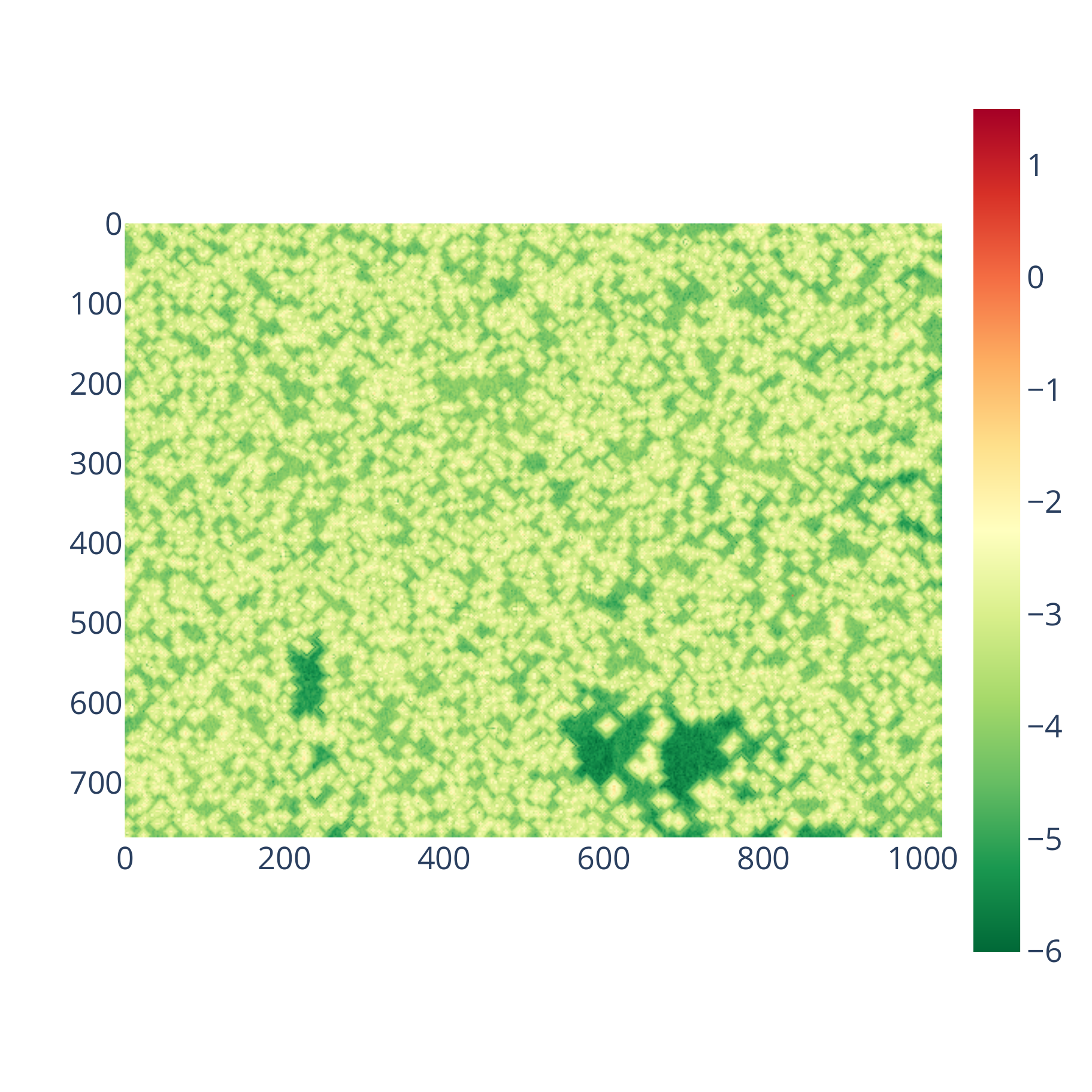}
    \caption{\centering\texttt{sepfir2d} standard deviation. (log)}
    \label{fig:bspline_bisplev_std}
\end{subfigure}
\begin{subfigure}{0.3\linewidth}
    \includegraphics[width=\linewidth]{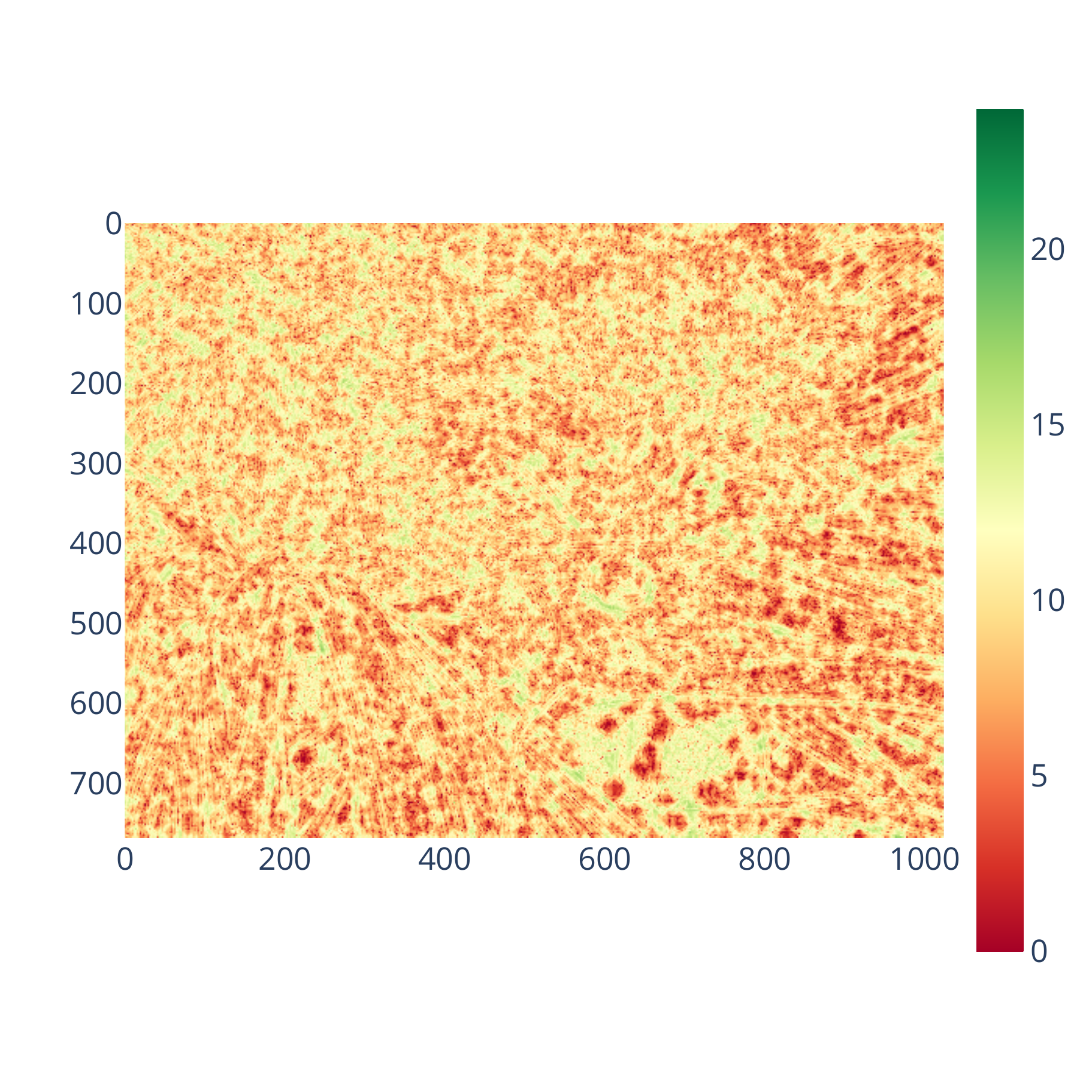}
    \caption{\centering\texttt{convol2d} significant bits. \newline \textcolor{white}{.}}
    \label{fig:bspline_convol2d_sig}
\end{subfigure}
\begin{subfigure}{0.3\linewidth}
    \includegraphics[width=\linewidth]{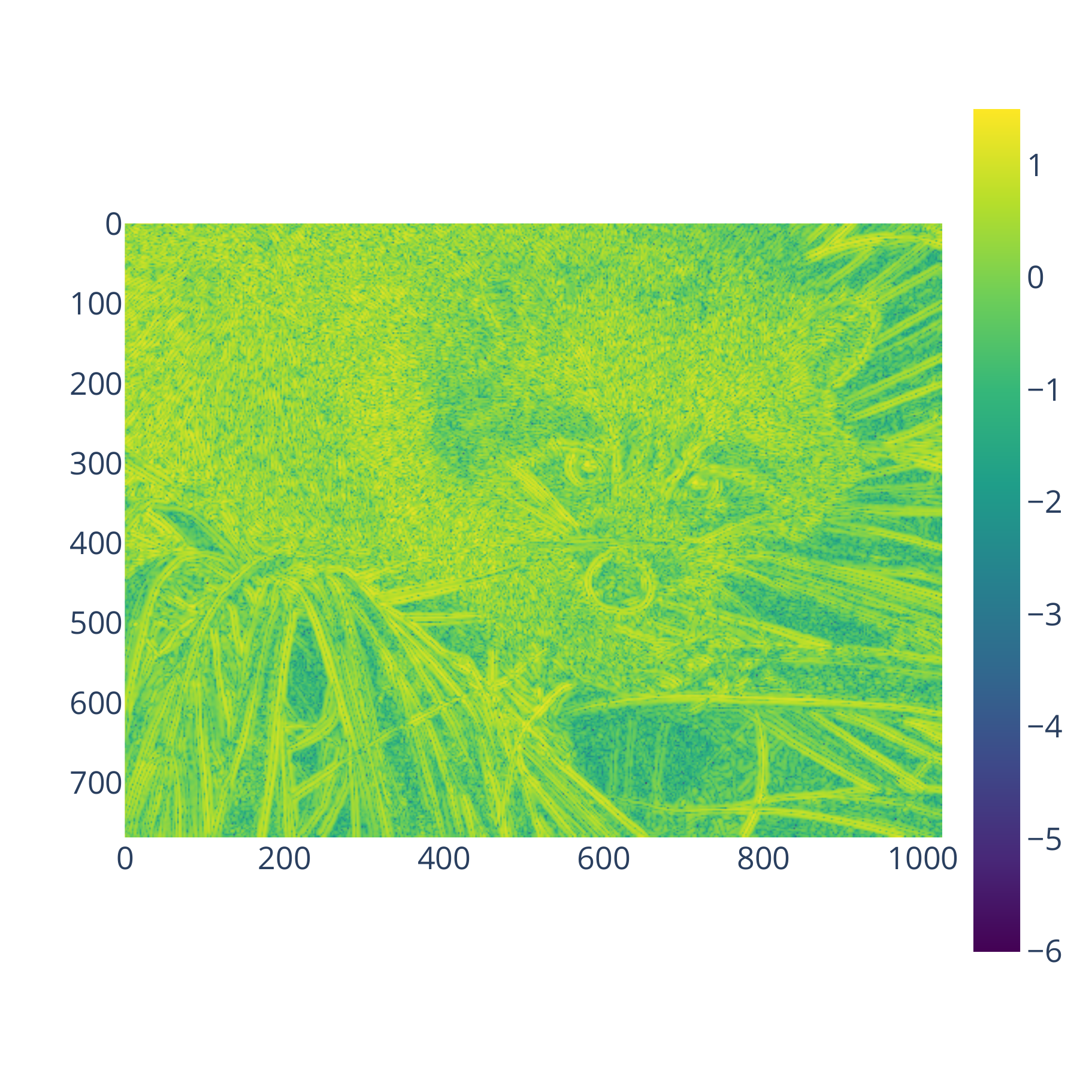}
    \caption{\centering\texttt{convol2d} mean. \newline (log)}
    \label{fig:bspline_convol2d_mean}
\end{subfigure}
\begin{subfigure}{0.3\linewidth}
    \includegraphics[width=\linewidth]{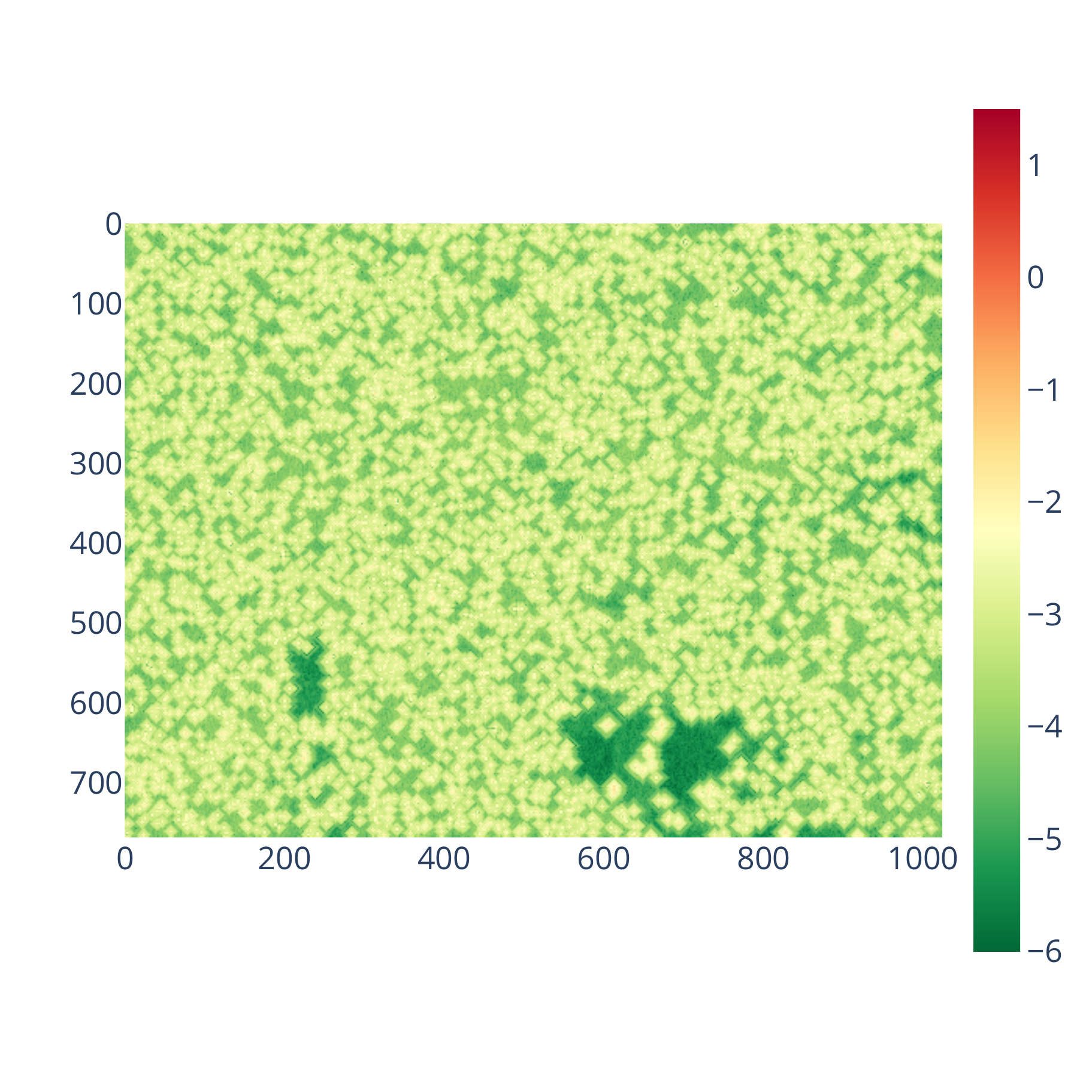}
    \caption{\centering\texttt{convol2d} standard deviation. (log)}
    \label{fig:bspline_convol2d_std}
\end{subfigure}
    \caption{\texttt{bspline} results within RR mode. \texttt{sepfir2d} and
 \texttt{convol2d} have a similar precision.
    }
    \label{fig:bspline_rr}
\end{figure}

\texttt{bspline} compares two edge filters: \texttt{bisplev} evaluates a bivariate B-spline and its derivative whereas \texttt{convolved2d} convolves two 2-dimensional arrays. Figure~\ref{fig:bspline_rr}
shows the significant bits, mean, and standard deviation 
maps for the two methods as well as the input image (Fig.~\ref{fig:bspline_original_image}). The middle row shows \texttt{sepfir2d} results and the bottom row shows \texttt{convol2d} results. Both methods exhibit similar precision, with 11 significant bits on average. Standard deviations maps on Figures~\ref{fig:bspline_bisplev_std} and~\ref{fig:bspline_convol2d_std} show that regions with low spatial frequencies correspond to regions with low standard deviation, similar to the FFT results. The SciPy tutorial mentions that \texttt{bisplev} is faster than \texttt{convol2d}. The comparable numerical precision observed in our experiments reinforce the use of \texttt{bisplev} over \texttt{convol2d} in this example.

\texttt{spline\_1D} computes the 1D spline interpolation for $f(x)=\sin(x)$ on points $x=\frac{2\pi k}{8}, k \in [0, 10]$. The spline interpolation uses \texttt{splrep} to build the spline representation and \texttt{splrev} to evaluate the spline at any point. It also computes the integral by using \texttt{splint} and the root finder \texttt{sproot}. Although the traces diverge, we can analyze the partial aggregation for the \texttt{splrep}, \texttt{splev} and \texttt{splint}. The three functions show results with an average precision of 51 bits with RR mode.
 
\begin{figure}
\begin{subfigure}{.3\textwidth}
    \centering
    \includegraphics[width=\linewidth]{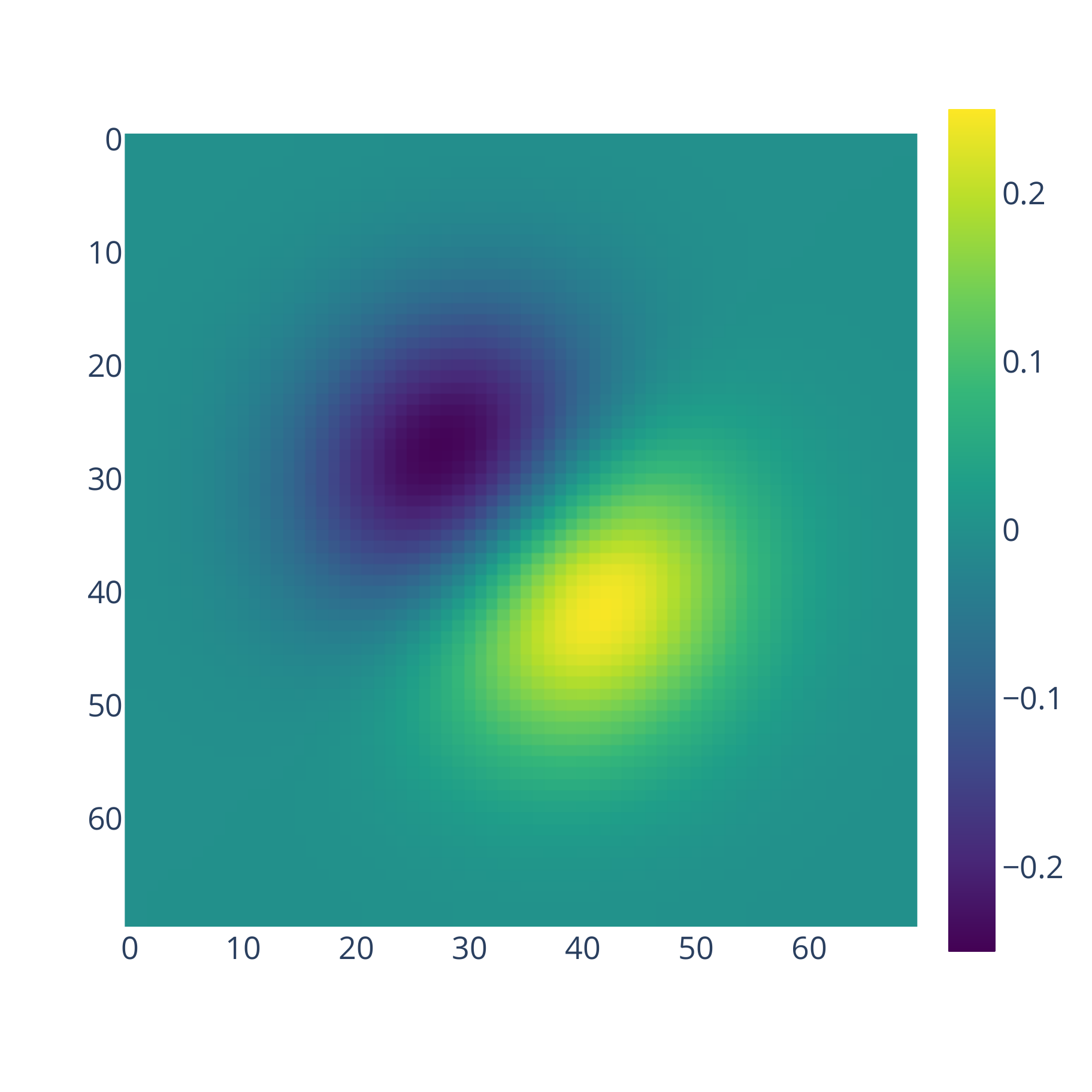}
    \caption{Mean value}
    \label{fig:bisplev_mean}
\end{subfigure}
\begin{subfigure}{.3\textwidth}
    \centering
    \includegraphics[width=\linewidth]{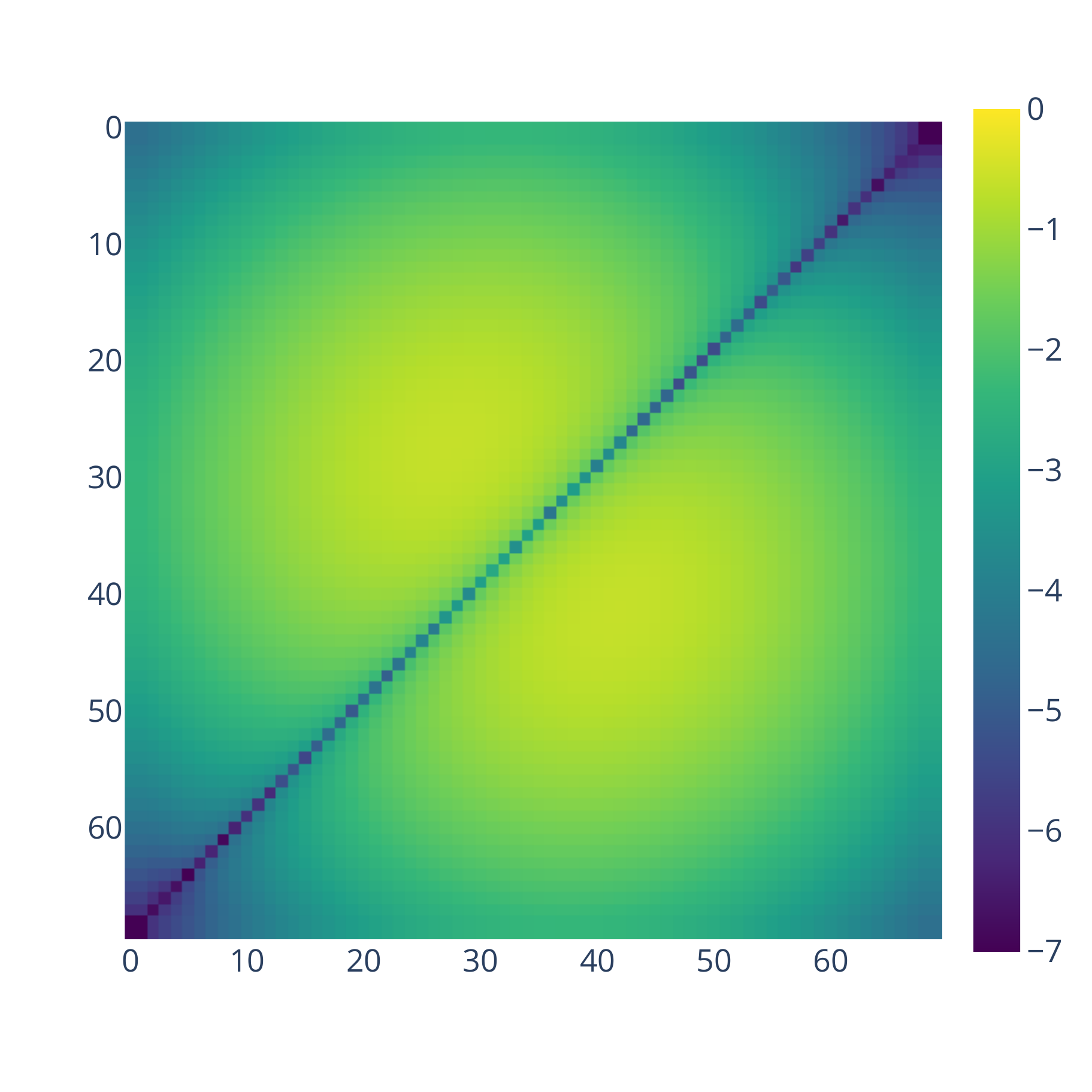}
    \caption{Absolute mean value (log)}
    \label{fig:bisplev_mean_log}
\end{subfigure}
\begin{subfigure}{.3\textwidth}
    \centering
    \includegraphics[width=\linewidth]{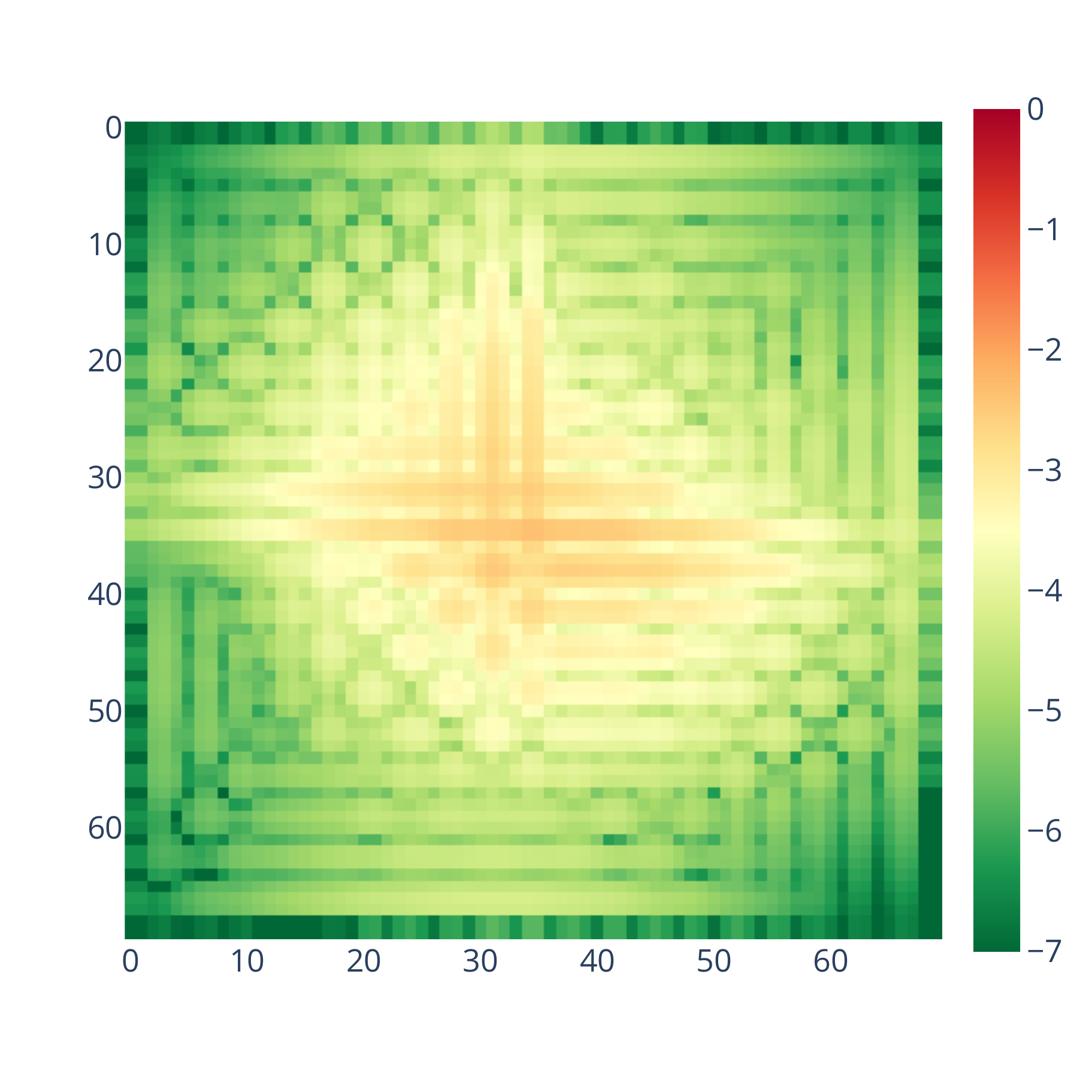}
    \caption{Standard deviation (log)}
    \label{fig:bisplev_std_log}
\end{subfigure}
    \caption{2D Spline interpolation results. Figure~\ref{fig:bisplev_mean} shows the mean result
    of the 2D-spline interpolation. Figures~\ref{fig:bisplev_mean_log} and~\ref{fig:bisplev_std_log}
    show respectively in the logarithmic scale the absolute mean value and the standard deviation 
    of the 5 samples run within RR mode. 
    }
    \label{fig:spline2d_rr}
\end{figure}

\texttt{spline\_2D} computes the 2D spline interpolation for function $z=(x+y)e^{-6(x^2+y^2)}$, with $(x,y) \in [-1,1]\times[-1,1]$ and a sampling of 21 points for each coordinate. \texttt{bisplrep} builds the spline representation with the $21 \times 21$ points of $z$ over 71 grid sampling points for $(x,y)$. Figure~\ref{fig:spline2d_rr} shows the results of the spline evaluation with \texttt{bisplev}.  We observe a significant loss of precision in places, which follows 
an interesting pattern that could be further investigated. 



\subsubsection{Optimization}

This package has eleven examples. Seven examples involve the same optimization problem 
solved with and without constraints using different methods.
One example optimizes a chemical problem under constraints.
The last three examples solve a different problem each.

\paragraph{Unconstrained minimization of multivariate scalar functions:}
Four Quasi-Newton-Raphson methods minimize function $f$ by using Newton's step: $x_{k+1} = x_{k} - B^{-1}(x_k)\nabla f(x_k)$, where $\nabla$ and $B$ are the gradient and the approximation of the Hessian matrix of $f$. The \texttt{Broyden-Fletcher-Goldfarb-Shanno}~\cite{BFGS} (BFGS) and \texttt{Newton-Conjugate-Gradient}~\cite{nocedal2006numerical} (NCG) are line search methods that approximate $H$ by adding two symmetric rank-one matrices and by using the Conjugate-Gradient method. The \texttt{Trust-Region Truncated Generalized Lanczos / Conjugate Gradient}~\cite{gould1999solving} (TR-Krylov), \texttt{Trust-Region Newton-Conjugate-Gradient} (TR-NCG), and \texttt{Trust-Region Nearly Exact}~\cite{nocedal2006numerical} (TR-Exact) are trust-region methods approximating $H$ by solving the trust-region subproblem restricted to a truncated Krylov subspace and by solving nonlinear equations for each quadratic subproblem. 
In addition, the \texttt{Nelder-Mead}~\cite{singer2009nelder} simplex-based method iteratively transforms a simplex until its vertices are getting closer to a local minima of the function. All methods are applied to the Rosenbrock function.

\paragraph{Constrained minimization of multivariate scalar functions:}
Two optimizers solve an optimization problem by taking into account constraints.
The \texttt{Sequential Least SQuares Programming} uses the SLSQP~\cite{kraft1988software} method to optimize the Rosenbrock function.
The \texttt{Least-squares minimization} uses the Trust Region Reflective algorithm~\cite{li1993centering}
to solve a fitting problem from an enzymatic reaction.

\paragraph{}
The SLSQP and unconstrained minimization methods minimize the \textbf{Rosenbrock} function of $N$ variables
that reaches its minimum (0) when $x_i=1,\; \forall i \leq N-1$:
\[f(x) = \sum_{i=1}^{N-1} 100(x_{i+1}-x^2_i)^2 + (1-x_i)^2\]

\begin{figure}
    \centering
    \includegraphics[width=\linewidth]{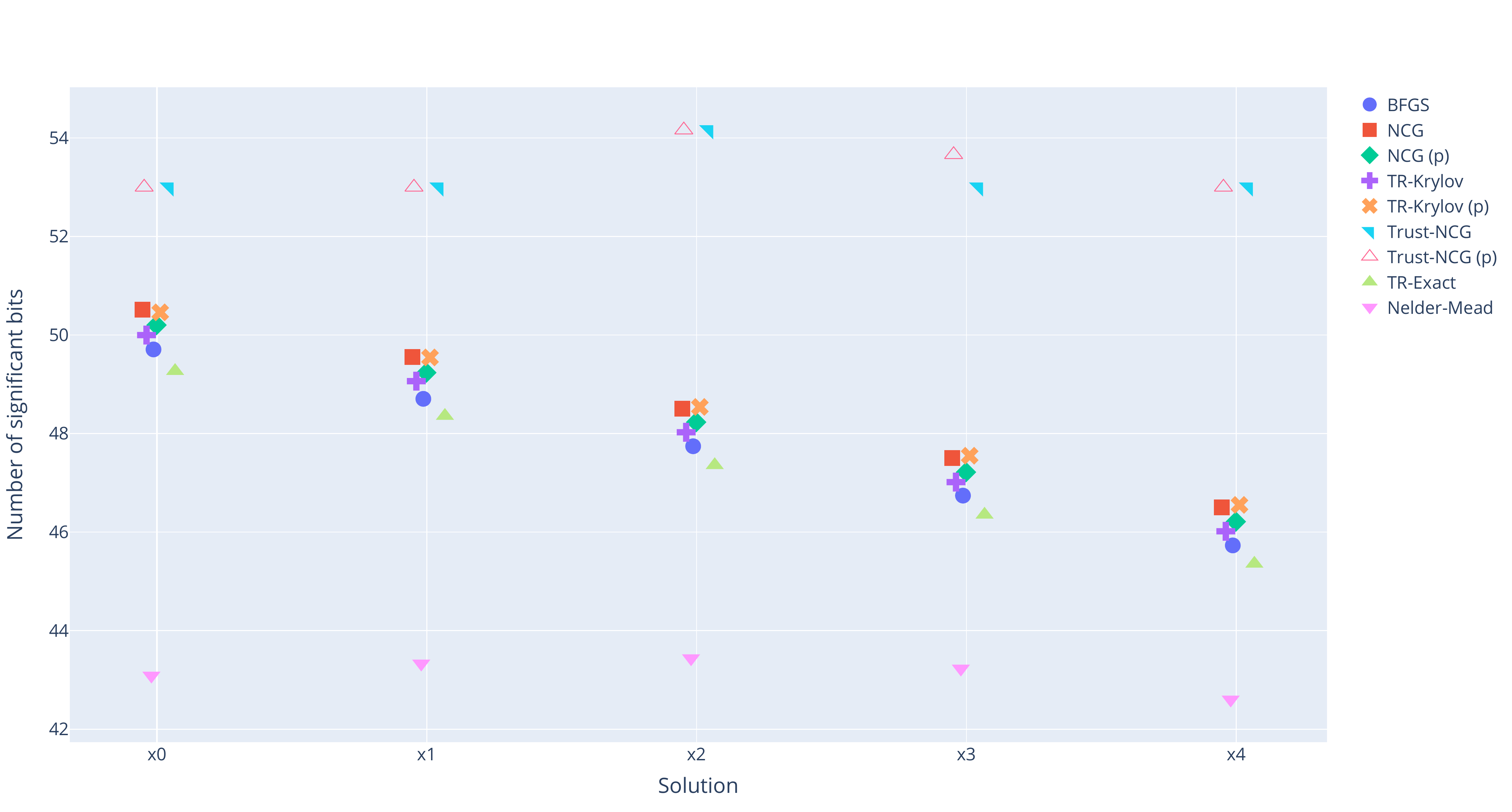}
    \caption{Comparison of results precision for different optimization solvers within RR mode.}
    \label{fig:unconstrained_optimization}
\end{figure}

Figure~\ref{fig:unconstrained_optimization} shows the number of significant bits of the solution for the different methods used. 
The $(p)$ notation for the NCG, Trust-Krylov, and Trust-NCG methods in the legend refers 
to the variant where the Hessian matrix is replaced by 
a function computing the matrix-vector product of the Hessian and an arbitrary vector, to save memory and time.
All the methods have a good precision ranging from 43 to 53 significant bits. The Nelder-Mead method is the least precise one with $\simeq$ 43 significant bits on average, while the Trust-Region-Newton-CG is the most precise one with $\simeq$ 53 significant bits, the highest precision achievable in double precision. The remaining methods have a similar precision. 

The SLSQP also minimizes the Rosenbrock function with the following constraint:
\begin{eqnarray*}
    x_0 + 2x_1 \leq 1 \\
    x_0^2 + x_1 \leq 1 \\
    x_0^2 - x_1 \leq 1 \\
    2x_0 + x_1 = 1 \\
    0 \leq x_0 \leq 1 \\
    -0.5 \leq x_1 \leq 2 
\end{eqnarray*}
The problem has a unique solution $[x_0, x_1] = [0.4149, 0.1701]$.
Results obtained with RR show a precision of 47 and 44 significant bits for the first and the second solution. 

The \texttt{least-square minimization} example solves a fitting problem from an enzymatic reaction~\cite{kowalik1968analysis} with 11 residuals defined as:
\begin{eqnarray*}
f_i(x) &=& \frac{x_0(u_i^2 + u_ix_1)}{u_i^2 + u_ix_2+x_3}-y_i,\; i=0,...,10 \\
&0& \leq x_j \leq 100,\; j=0,..,3
\end{eqnarray*}
where where $y_i$ are measurement values, $u_i$ are values of the independent variable, and
$x_i$ the unknown.
Results within RR show a precision of 51, 46, 47, and 48 significant bits for each component of the solution respectively.

\paragraph{Root finding:}

The \texttt{root\_*} examples use three algorithms to 
find the root of non-linear equations. \texttt{root\_finding}
uses the hybrid Powell method to solve a single-variable transcendental equation and 
the Levenberg-Marquardt method for a set of non-linear equations. 
\texttt{root\_finding\_large} and \texttt{root\_finding\_large\_preconditioned} examples use 
the Krylov method to approximate the inverse Jacbian with and without the help of a preconditioner. 
They solve an integrodifferential equation on a square
 $(x,y) \in [0,1] \times [0,1]$, $P(x,1)=1$, and $P=0$ elsewhere on the boundary.
The function $P$ is discretized with a Cartesian grid $P_{n,n}=P(nh,nh)$, with $n=75$, and its derivatives are approximated by $\partial_x^2P(x,y) \simeq (P(x+h,y) - 2P(x,y) + P(x-h,y))/h^2$.

\begin{figure}
    \centering
    \begin{subfigure}{0.45\linewidth}
    \includegraphics[width=\linewidth]{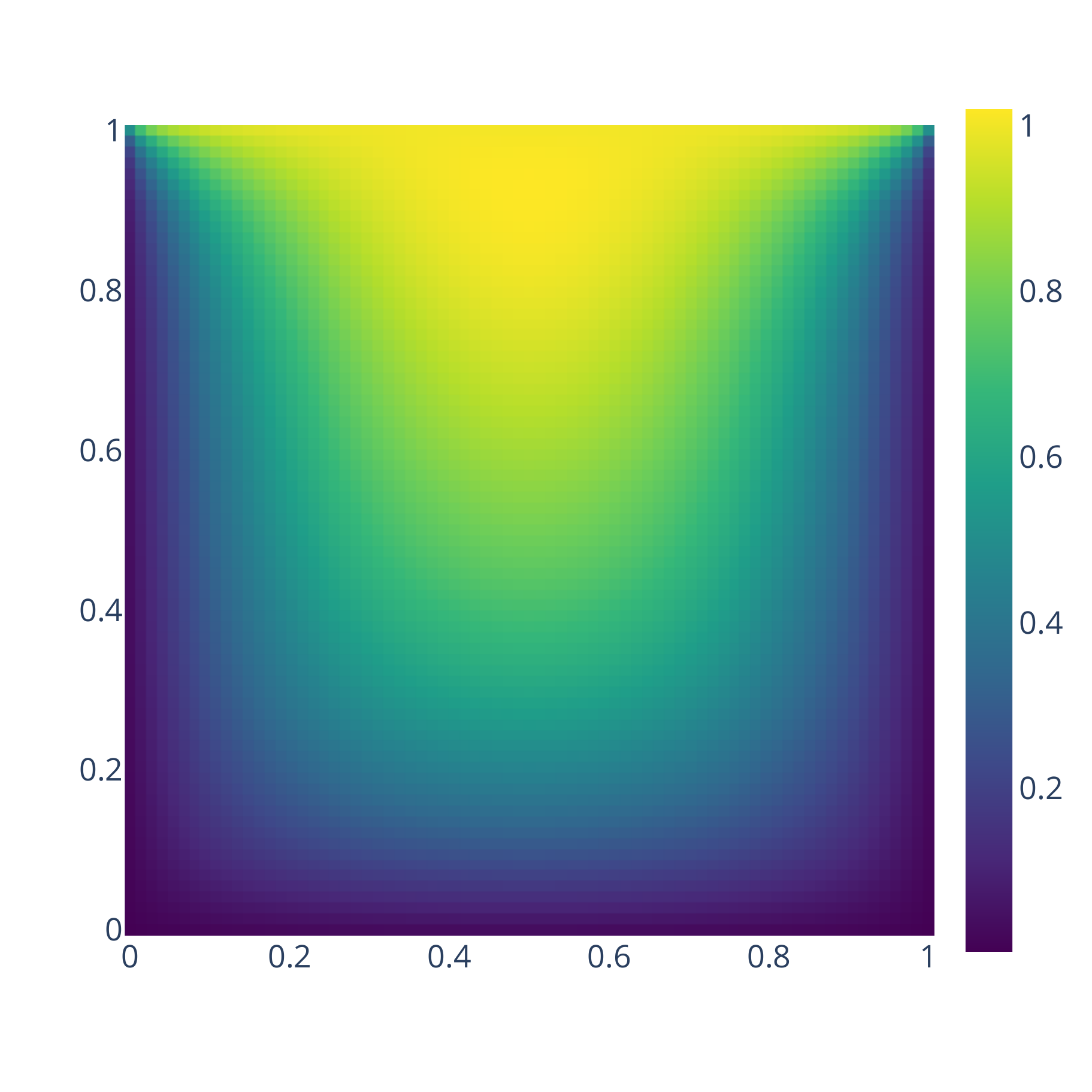}
    \caption{Mean solution (RR)}
    \label{fig:mean_solution_rr}
    \end{subfigure}
    \begin{subfigure}{0.45\linewidth}
    \includegraphics[width=\linewidth]{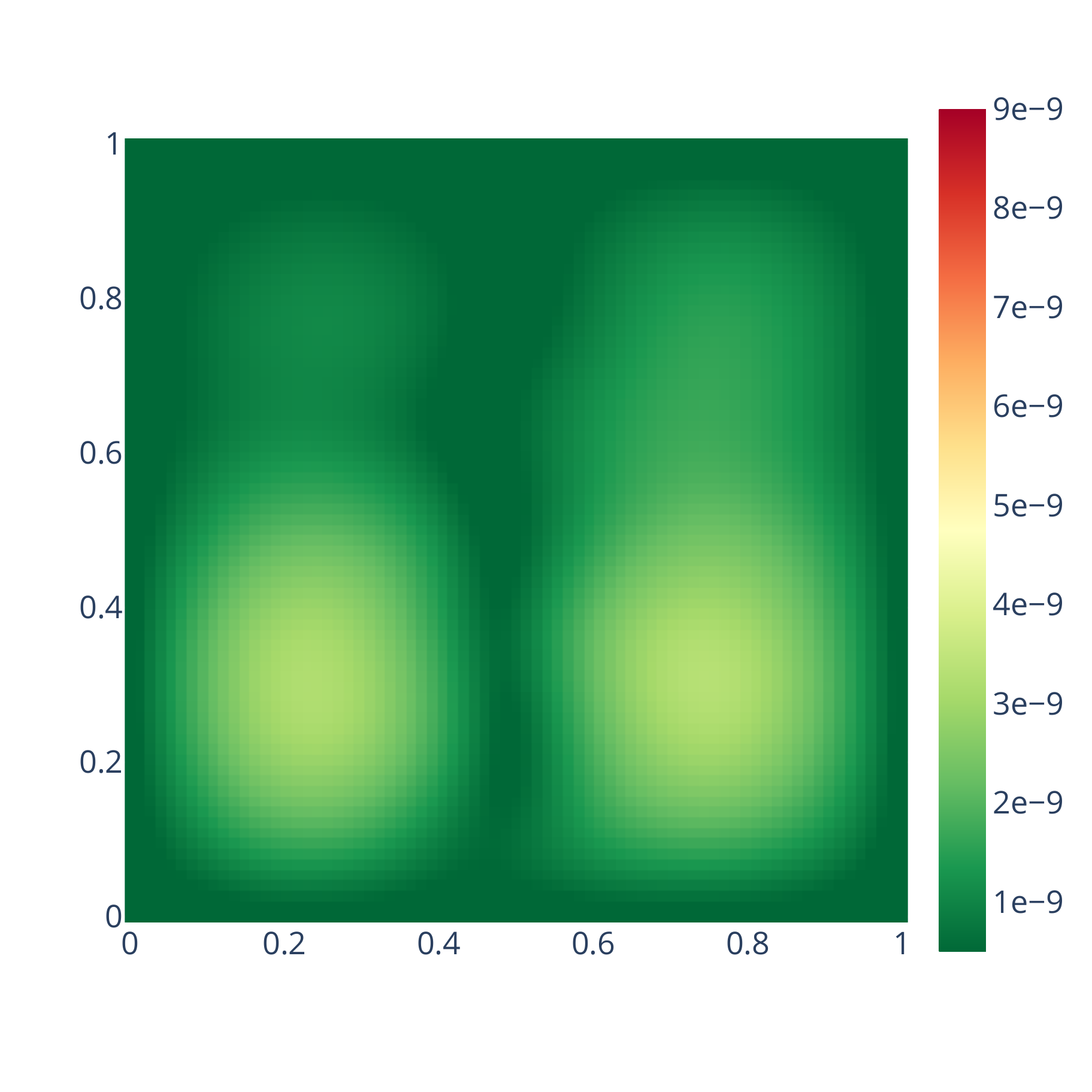}
    \caption{Standard deviation solution (RR)}
    \label{fig:stdev_rr}
    \end{subfigure}
    
    \begin{subfigure}{0.45\linewidth}
    \includegraphics[width=\linewidth]{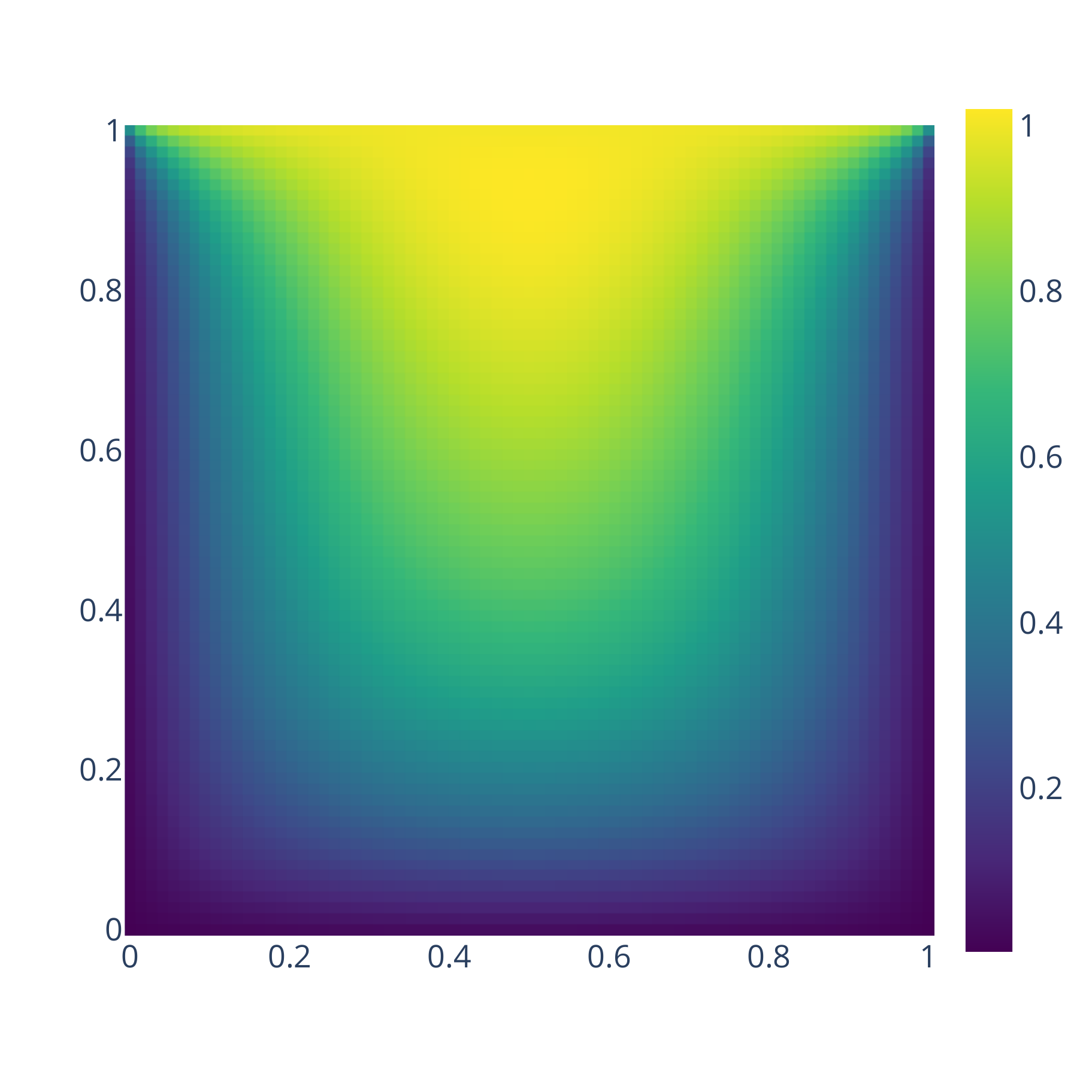}
    \caption{Mean solution (Full MCA)}
    \label{fig:mean_solution_mca}
    \end{subfigure}
    \begin{subfigure}{0.45\linewidth}
    \includegraphics[width=\linewidth]{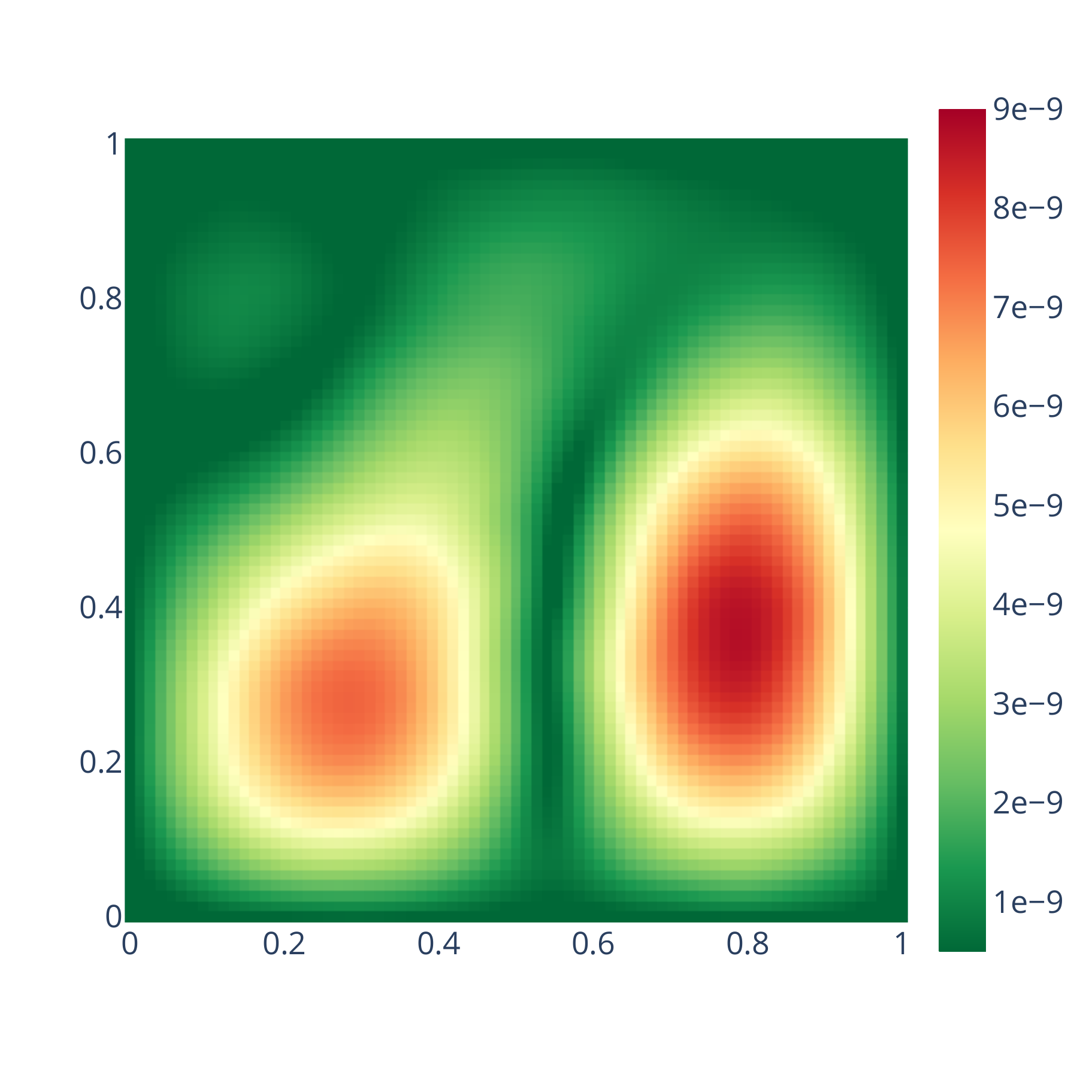}    
    \caption{Standard deviation (Full MCA)}
    \label{fig:stdev_mca}
    \end{subfigure}
    \caption{Solution of \texttt{root\_finding\_large} example. 
    The standard deviation has two maximum for both RR and Full MCA modes. The standard deviation remains lower than
    the stopping criterion threshold fixed at $6.10^{-6}$. }
    \label{fig:root_finding_large}
\end{figure}

All the examples diverge even within RR, which is not surprising since root-finding methods can easily take a few extra steps to find the root depending on the starting point. Nevertheless, for the \texttt{root\_finding\_large} example, 3 traces among the 5 samples can be merged for both RR and MCA modes.
Figure~\ref{fig:root_finding_large} shows the mean and standard deviation of the result for these 3 samples. We can see that a similar solution was found for RR and Full MCA. Full MCA leads to a lower precision than RR although both standard deviation maps remain in the same order of magnitude. 
Finally, it is interesting to note the impact of preconditioning on the result numerical quality: as Table~\ref{tab:pytracer_results_summary} shows, the precision doubles between \texttt{root\_finding\_large} and \texttt{root\_finding\_large\_preconditioned}.

\subsection{Scikit learn}
\label{sec:sklearn_tests}

Scikit-learn offers a well-supplied and documented list of examples 
that facilitates its use with Pytracer.
Among the several available examples, we choose the following representative set
listed in Table~\ref{tab:pytracer_results_summary}. The following paragraphs 
focus on the most interesting stability results observed across all experiments.
\begin{figure}
    \centering
    \includegraphics[width=0.5\linewidth]{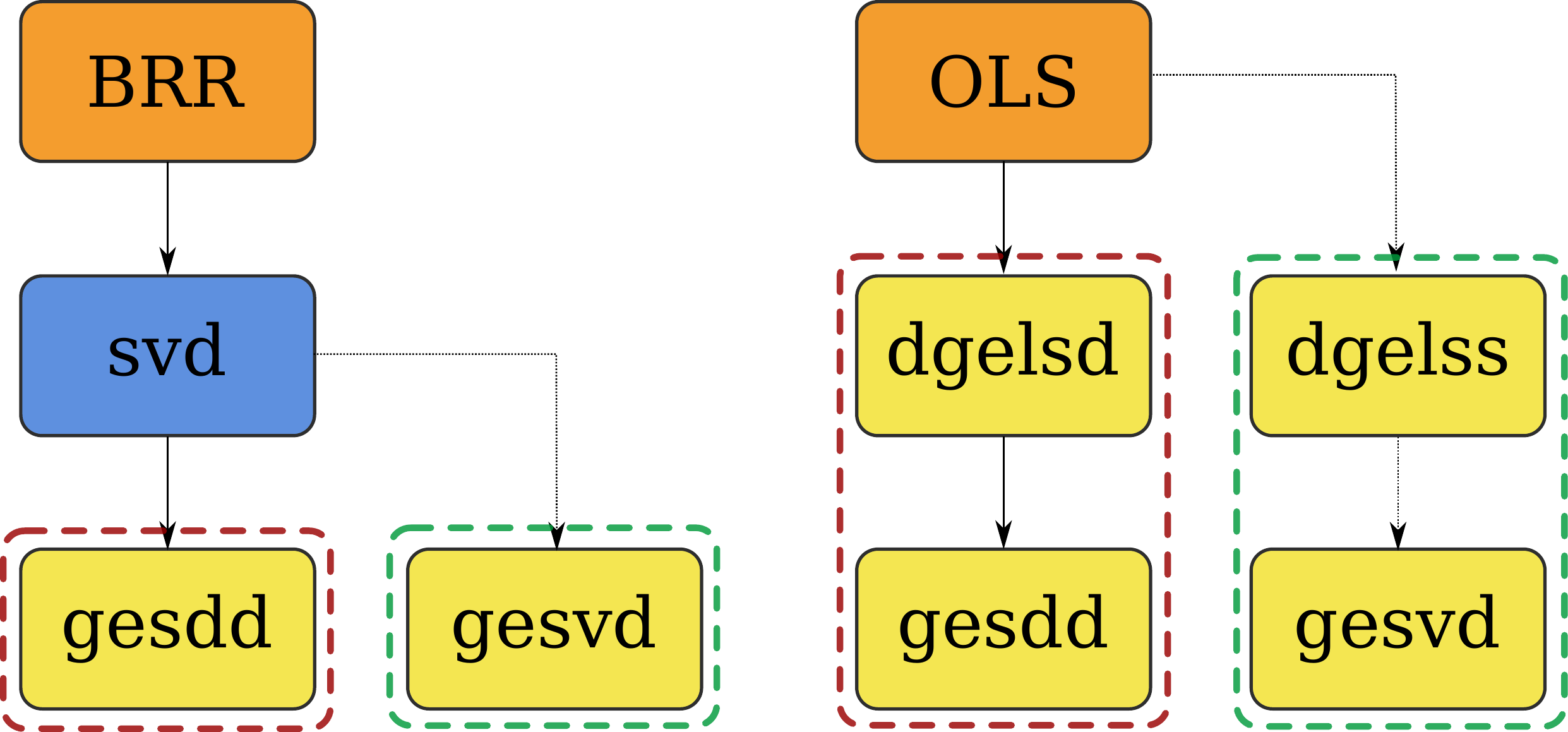}
    \caption{Call paths of the Bayesian Ridge Regression (BRR) and 
    Ordinary Least Square (OLS) function. Color represents the calling library:
    orange for scikit-learn, blue for SciPy and yellow for LAPACK.
    Dashed color represents the original path in red and the alternative one in green.
    Figure~\ref{fig:brr_svd_sig} shows the original method  leads to numerical instabilities while the alternative one converges.
    }
    \label{fig:call_path_brr}
\end{figure}

\paragraph{Bayesian Ridge Regression}

\begin{figure}
    \centering
    \includegraphics[width=\linewidth]{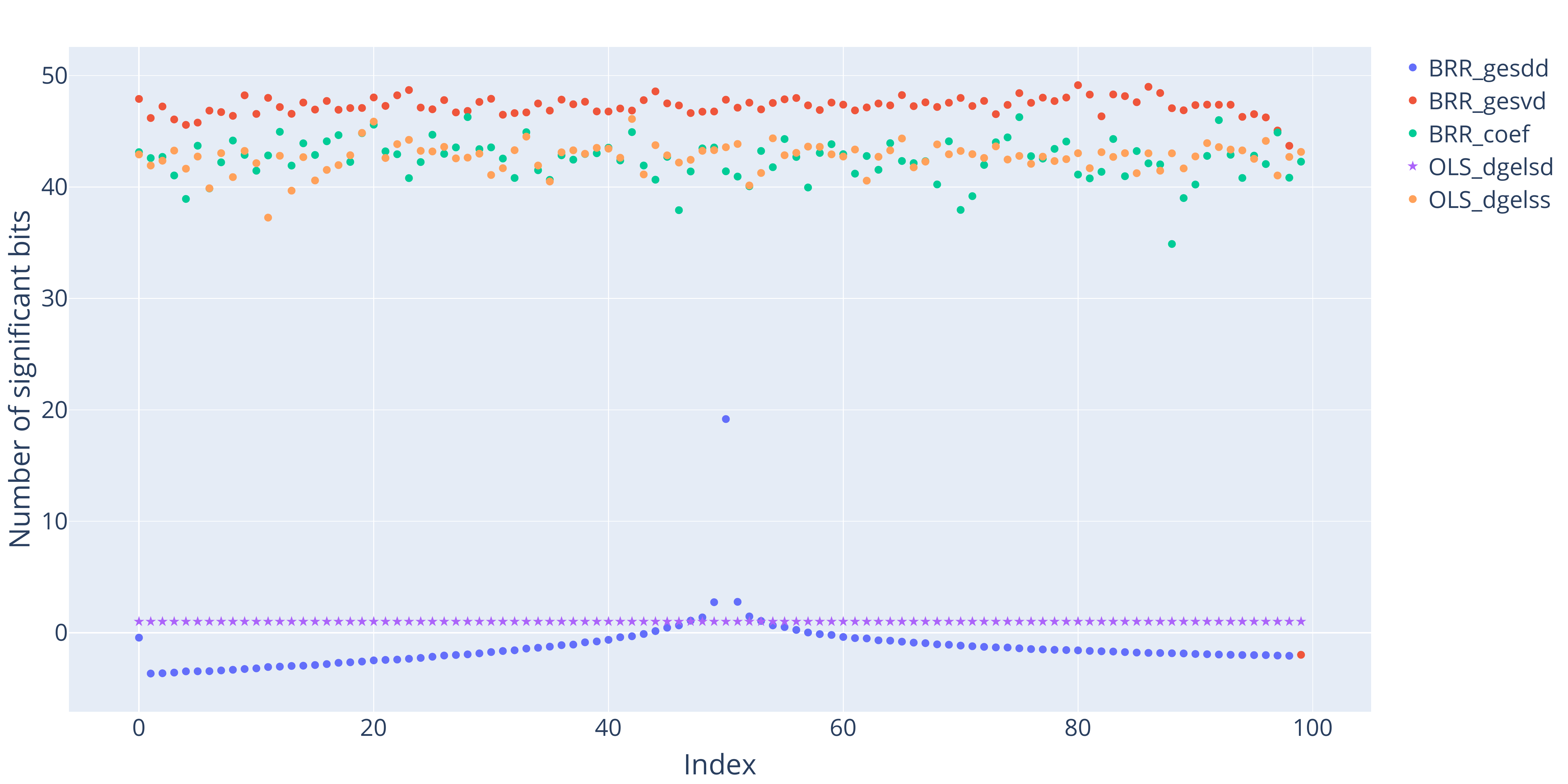}
    \caption{Precision of Bayesian Ridge Regression coefficients in RR mode.
     \texttt{BRR\_*} show the \texttt{BayesianRidge} results
    using the \texttt{gesdd} and \texttt{gesvd} methods. 
    \texttt{OLS\_*} show the \texttt{LinearRegression} results
    with the \texttt{dgelsd} and \texttt{dgelss} methods.
   The Divide \& Conquer method 
    (\texttt{gesdd} and  \texttt{dgelsd})
    has low precision
    for \textit{BRR} and results in \texttt{NaN} values (star points) for \textit{OLS}. By switching to the \texttt{gesvd} method, 
    results have a precision of 42 bits on average.
    }
    \label{fig:brr_svd_sig}
\end{figure}

This example compares a Bayesian Ridge Regression (BRR) to the Ordinary Least Squares (OLS) estimator on a synthetic dataset and for one-dimensional regression using polynomial feature expansion. Although the example does not raise a runtime error, the regression coefficients computed from the fitting are 
non-significant. \pytracer traces reveal that the SVD solver used is the root cause for this error. 
Figure~\ref{fig:call_path_brr} shows the two call paths for BRR and OLS. 
The LAPACK library has two main methods for computing the SVD: \texttt{gesdd} and \texttt{gesvd}.
The former uses a Divide \& Conquer approach, while the latter uses a QR decomposition. 
While both methods are expected to have the same accuracy~\cite{nakatsukasa2013stable}, Figure~\ref{fig:brr_svd_sig} shows that \texttt{gesdd} is totally imprecise, with an average number of significant digits of 0 for BRR and even \texttt{Not-a-Number} (\texttt{NaN}) values for OLS. 
On the same figure, we can see that 
using the \texttt{gesvd} (dgelsd) method instead of the \texttt{gesdd} (dgelss) one, 
the SVD converges for RR and Full MCA, with a number of significant bits of 44 on average. 
This observation supports the on-going discussion among LAPACK developers on the instability of \texttt{gesdd} (see  \href{https://github.com/Reference-LAPACK/lapack/issues/316}{github.com/Reference-LAPACK/lapack/issues/316}). 


\paragraph{Face Recognition}

The face recognition example uses eigenfaces (eigenvectors) and Support Vector Machines (SVMs) to classify faces from the Labeled Faces in the Wild dataset. This example uses a Principal Component Analysis (PCA) to extract the eigenfaces and then trains an SVM classifier to predict the name associated with a face. This example is the only one to raise an error with RR due to the non-convergence of the Singular Value Decomposition (SVD) performed in the PCA.

\begin{figure}
    \centering
    \begin{subfigure}{0.3\linewidth}
    \includegraphics[width=\linewidth]{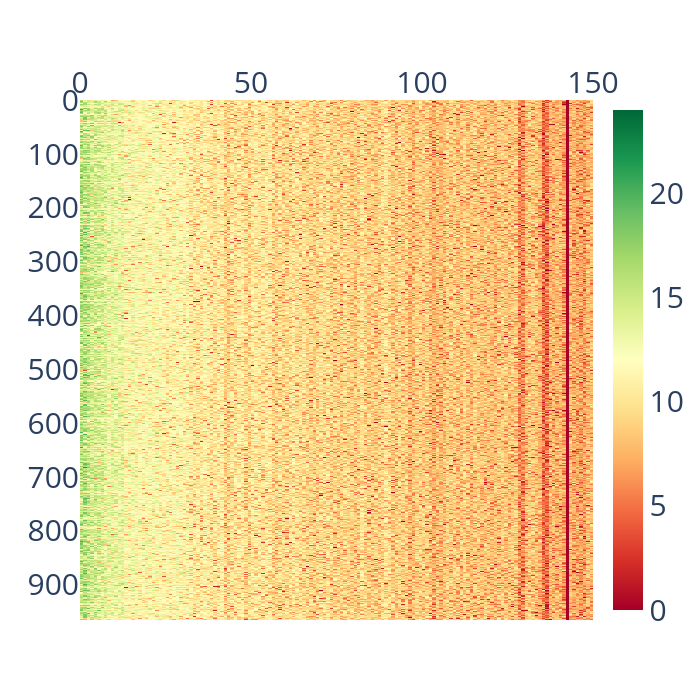}
    \caption{Left matrix $U$ of the SVD decomposition}
    \label{fig:randomized_svd_U}
    \end{subfigure}
    \begin{subfigure}{0.3\linewidth}
    \includegraphics[width=\linewidth]{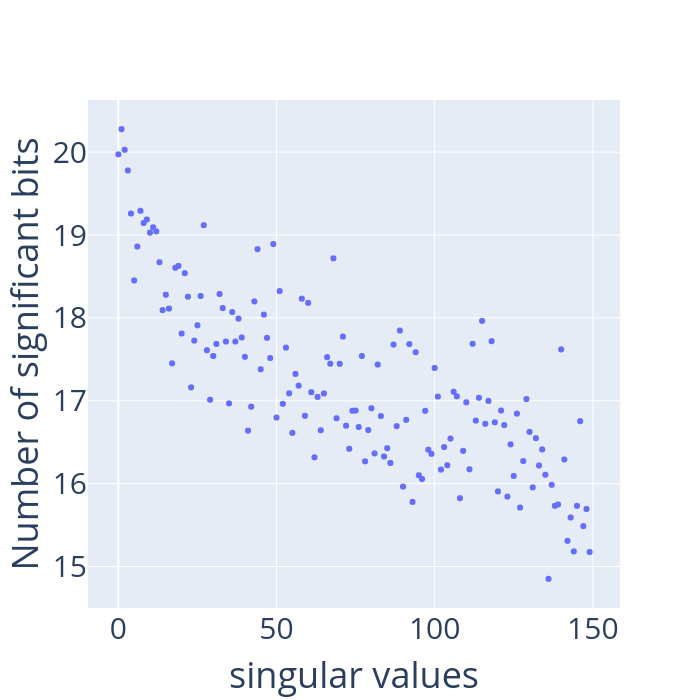}
    \caption{Singular values $\Sigma$ of the SVD decomposition}
    \label{fig:randomized_svd_S}
    \end{subfigure}
    \begin{subfigure}{0.3\linewidth}
    \includegraphics[width=\linewidth]{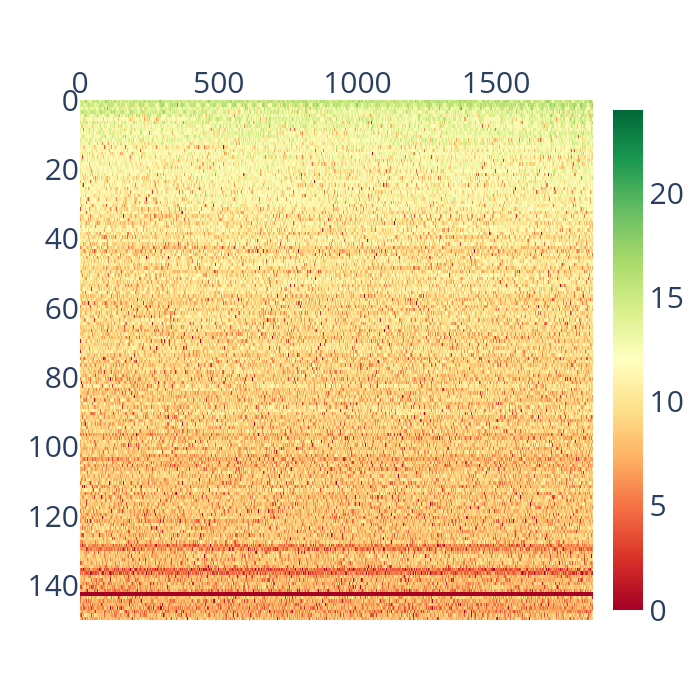}
    \caption{Right matrix $V$ of the SVD decomposition}
    \label{fig:randomized_svd_V}
    \end{subfigure}
    \caption{Precision of the \texttt{randomized\_svd} function in the 
    \texttt{face\_recognition} example using the \texttt{gesvd} SVD method and RR.
    Figures~\ref{fig:randomized_svd_U},~\ref{fig:randomized_svd_S}, and~\ref{fig:randomized_svd_V}
    show respectively the values $U$, $\Sigma$ and $V$.
     Numerical precision is lower for smaller singular values and this loss of precision translates to singular vectors. It would be interesting to investigate the use of numerical precision for dimensionality reduction.
    }
    \label{fig:face_recognition_svd}
\end{figure}

In this example, the PCA uses the Randomized SVD~\cite{halko2011finding} method that first computes a low-rank approximation of the input matrix using a stochastic algorithm and then uses this approximation as input for the SVD. The example raises an error during the computation of the SVD.
The \texttt{randomized\_svd} scikit-learn function uses the Divide \& Conquer SVD method \texttt{gesdd} mentioned previously.
Replacing \texttt{gesdd} for \texttt{gesvd} solves the SVD convergence issue. Figure~\ref{fig:randomized_svd_S} shows the precision of the eigenvalues computed with gesvd, which varies between 15 and 20 significant bits out of the 24 available in single precision. The classifier metrics (precision, recall, F1-score, and support) are identical to the IEEE execution for all MCA samples.

In conclusion, \pytracer and the MCA model were helpful to detect the instability of the \texttt{gesdd} method mentioned in several GitHub issues, including the LAPACK repository. Furthermore, it is worth mentioning that a Pull Request has recently been submitted to let the user select the SVD method and avoid this kind of issue. Thus, \pytracer can detect instabilities that affect fundamental scientific computing components such as the LAPACK library.

\paragraph{Separating hyperplane}

\begin{figure}
    \centering
    \begin{subfigure}{.3\linewidth}
    \caption{Separating hyperplane}
    \includegraphics[width=\linewidth]{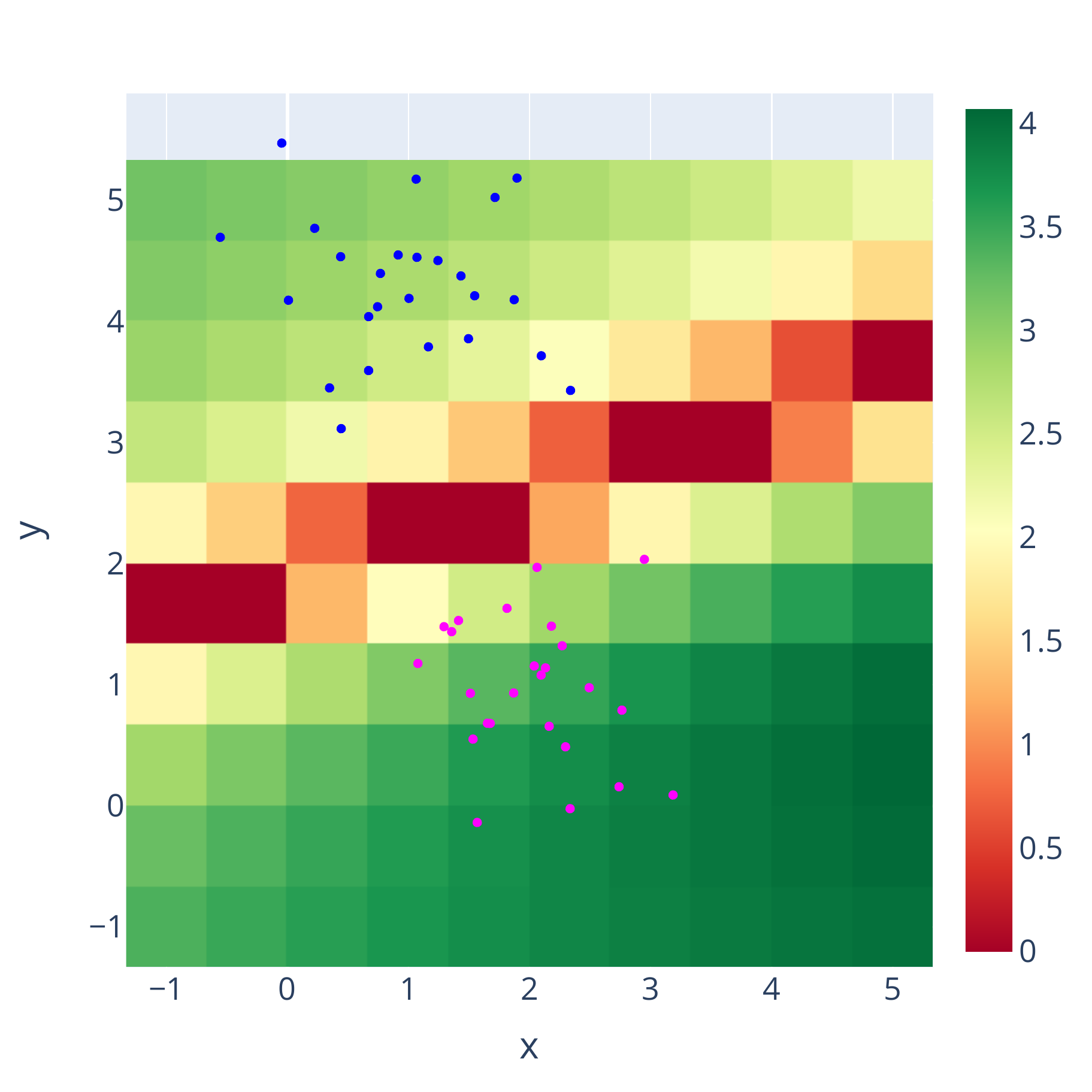}
    \label{fig:hyperplan_sig_zoom}
    \end{subfigure} \\
    \begin{subfigure}{.3\linewidth}
    \caption{SVM (non weighted)}
    \includegraphics[width=\linewidth]{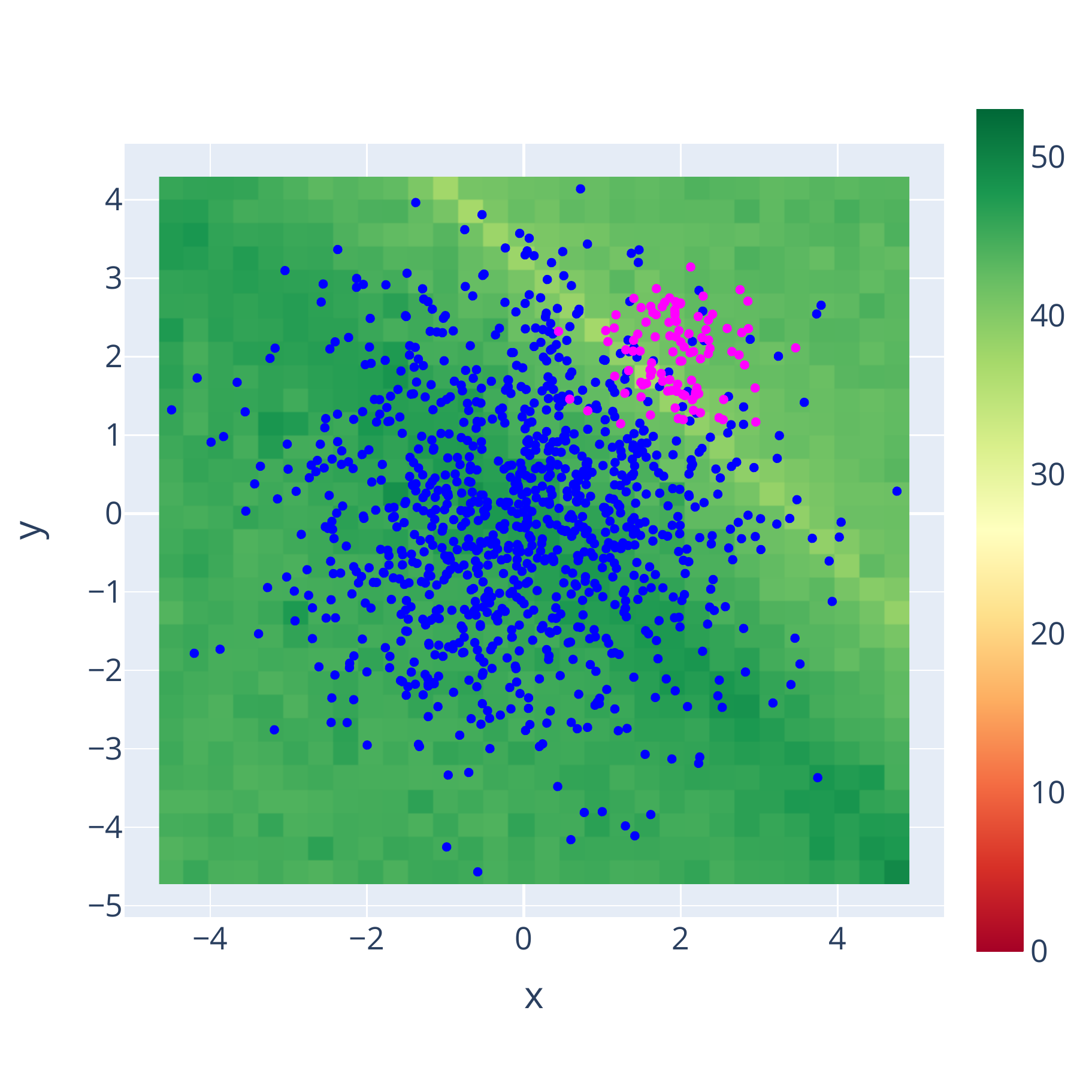}
    \label{fig:SVM_nw_sig}
    \end{subfigure}
    \begin{subfigure}{.3\linewidth}
    \caption{SVM (weighted)}
    \includegraphics[width=\linewidth]{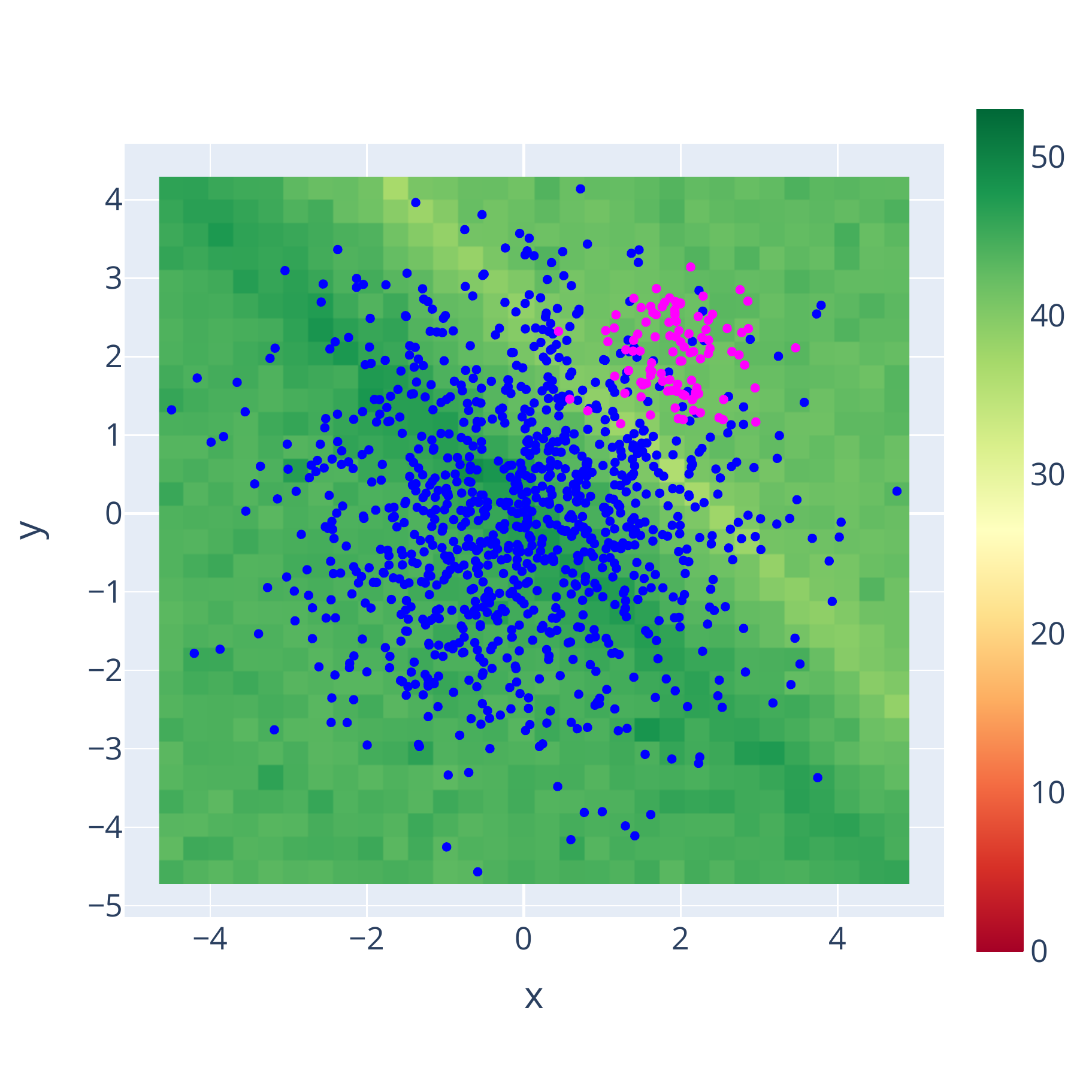}
    \label{fig:SVM_w_sig}
    \end{subfigure} \\
    \begin{subfigure}{.3\linewidth}
    \caption{SGD weighted (non-weighted)}
    \includegraphics[width=\linewidth]{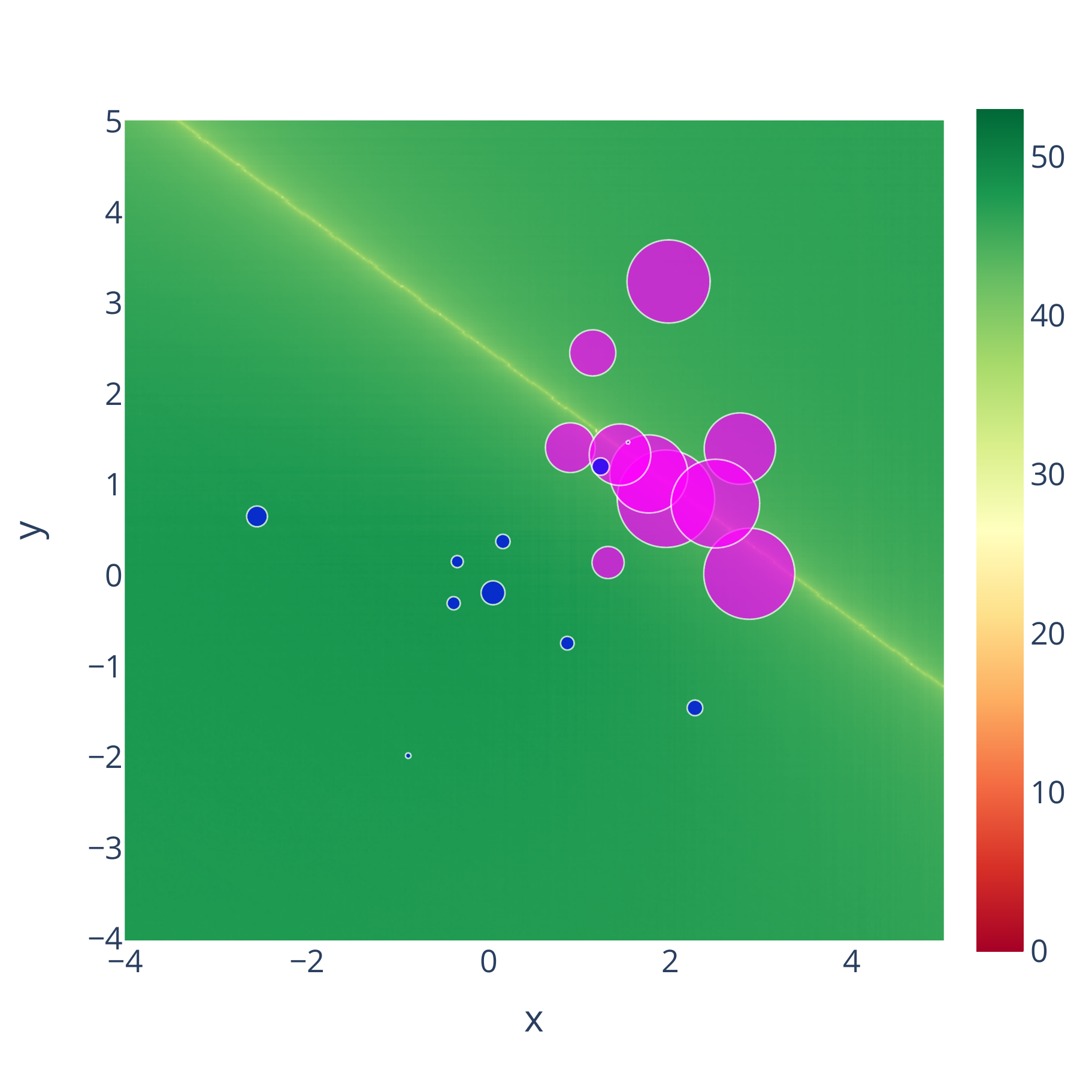}
    \label{fig:weighted_nw_sig}
    \end{subfigure}
    \begin{subfigure}{.3\linewidth}
    \caption{SGD weighted (weighted)}
    \includegraphics[width=\linewidth]{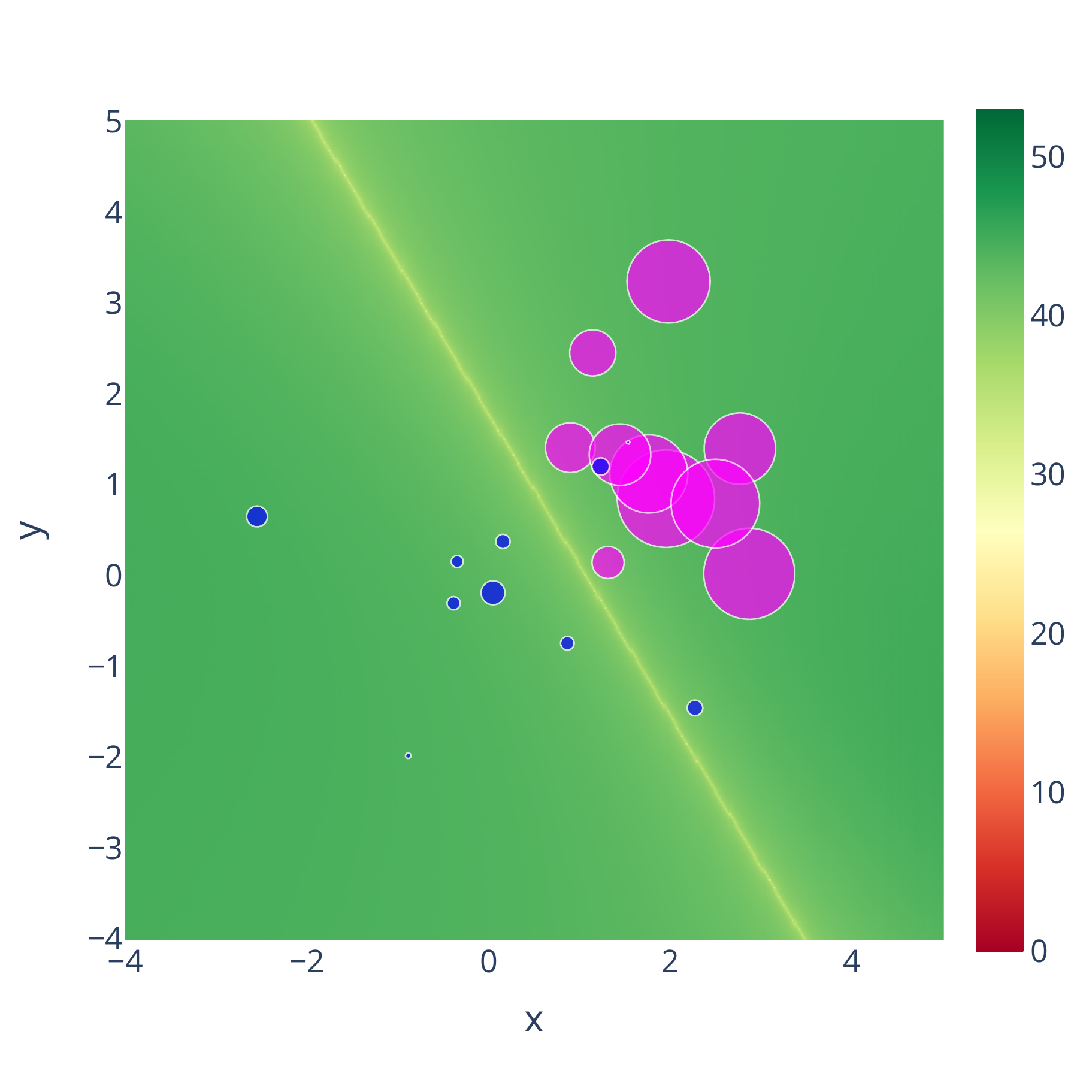}
    \label{fig:weighted_w_sig}
    \end{subfigure}
    \caption{
    Three examples of separating hyperplane prediction. All the figures show the number of significant bits of the prediction in RR mode. The separating hyperplane corresponds to a lower-precsion region in each figure. Using a large discretization step amplifies
    this instability.
    }
\label{fig:separating_hyperplan}
\end{figure}

This example uses a linear Support Vector Machine (SVM) classifier trained using the Stochastic Gradient Descent (SGD) method to find the maximum separating hyperplane between sets of 2D points.
The example trains the SVM with 50 random points separated in 2 clusters with hyperparameter $\alpha=0.01$ and using the hinge loss function. Once the model is trained, the example generates a $[-1,5]\times[-1,5]$ Cartesian grid
discretized with 100 points and predicts the class label for each point with integer coordinates. Figure~\ref{fig:hyperplan_sig_zoom} shows the number of significant bits 
obtained with RR for each point of the grid. 
We can see that yellow regions 
close to the separating hyperplane are highly unstable with a number of significant bits below 0, meaning that
the classifier is unsure about the predicted class. 


\paragraph{SVM}
This example uses a plain Support Vector Classification~\cite{Platt99probabilisticoutputs} (SVC) 
to find the best hyperplane separating an unbalanced dataset with two classes of 1000 and 100 points.
This example tests two SVC configurations using a linear kernel with and without a class weight parameter.
The SVC predicts the class label for each coordinate over a $[-5,5] \times [-5,5]$ grid discretized
with 900 points.
Figures~\ref{fig:SVM_nw_sig} and~\ref{fig:SVM_w_sig} show the result within RR of the classification with
and without taking into account the unbalancing.
We can see that the number of significant bits is lower on the separating line and
that taking into account class imbalance allows for a better separation.

\paragraph{SGD weighted samples}
This example uses an SGD to separate weighted points. It divides the training set into two classes of 10 points with a bigger weight for the second class. The SGD is trained with hyperparameters $\alpha=0.01$ and a maximum number of iterations to 100. The prediction is made on a $[-4,5]\times[-4,5]$ grid discretized with 250 000 points. Figures~\ref{fig:weighted_nw_sig} and~\ref{fig:weighted_w_sig} show the number of significant bits 
of prediction within RR along with dataset points in blue and violet.
We can see that the separating plane corresponds to a region of lower precision, similar to the previous examples. 




\subsection{Overhead evaluation}
\begin{figure}
    \centering
    \includegraphics[width=\linewidth]{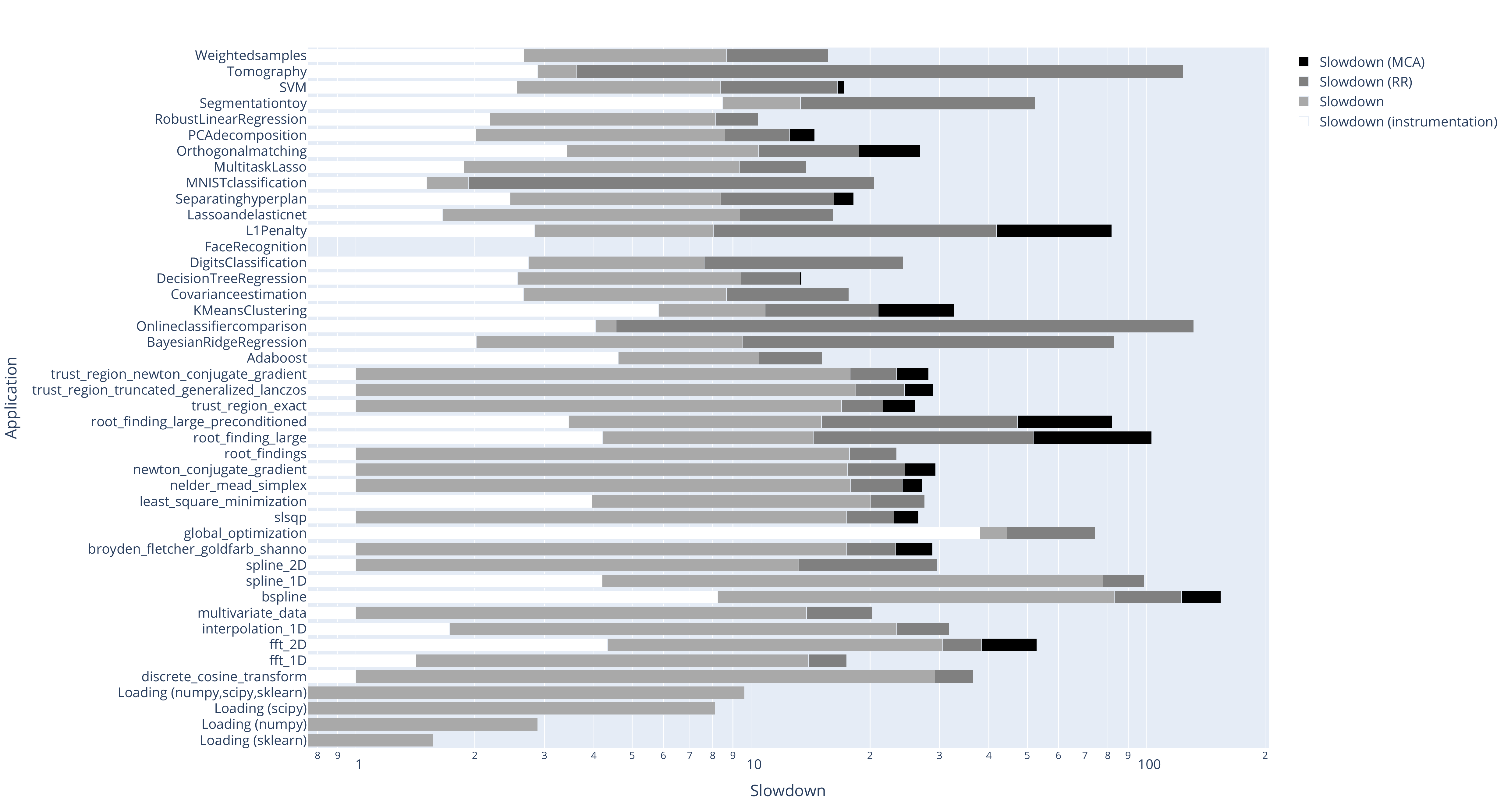}
    \caption{
    \pytracer tracing overheads. \textit{Slowdown} represents the slowdown of \pytracer without fuzzy measured as the ratio between the actual execution time (without \pytracer) and the execution time using \pytracer. 
    \textit{Slowdown (Instrumentation)} show the actual instrumentation cost computed by subtracting the initialization step.
    The \textit{Slowdown (RR)} and \textit{Slowdown (MCA)} are the slowdown of \pytracer 
    using the fuzzy environment with RR and Full MCA modes.
    The first four examples measure the initialization steps depending on the modules instrumented. 
    }
    \label{fig:performance_tracing}
\end{figure}

\pytracer's tracing overhead consists of (1) module instrumentation time (initialization overhead) and (2) data write time during the execution (runtime overhead). The initialization overhead is proportional to the number of functions and classes traced. For example, we measured an average initialization time of 8 seconds during our experiments with  NumPy, SciPy, and Scikit-learn. This overhead can be reduced by excluding intensively used functions such as \texttt{numpy.array} from the instrumentation.
Figure~\ref{fig:performance_tracing} summarizes \pytracer's tracing overhead for SciPy and scikit-learn examples. 
\textit{Slowdown} represents the slowdown of \pytracer without fuzzy measured as the ratio between the actual execution time (without \pytracer) and the execution time using \pytracer.
The \textit{Slowdown RR} and \textit{Slowdown MCA} show the slowdown within the fuzzy environment. 
Without fuzzy, \pytracer introduces an average slowdown of $17\times$ high for the examples running less than 8 seconds because \pytracer spends most of its time in the initialization step. Since this step is a constant factor of the number of functions instrumented, we also measured the instrumentation cost itself (\textit{Slowdown (Instrumentation)}) calculated by subtracting the initialization time to \textit{Slowdown}. We then obtain an average slowdown of $3.5\times$.
Finally, the fuzzy instrumentation increases the average slowdown to $37\times$ for RR and $43\times$ for Full MCA.


\begin{figure}
    \centering
    \includegraphics[width=\linewidth]{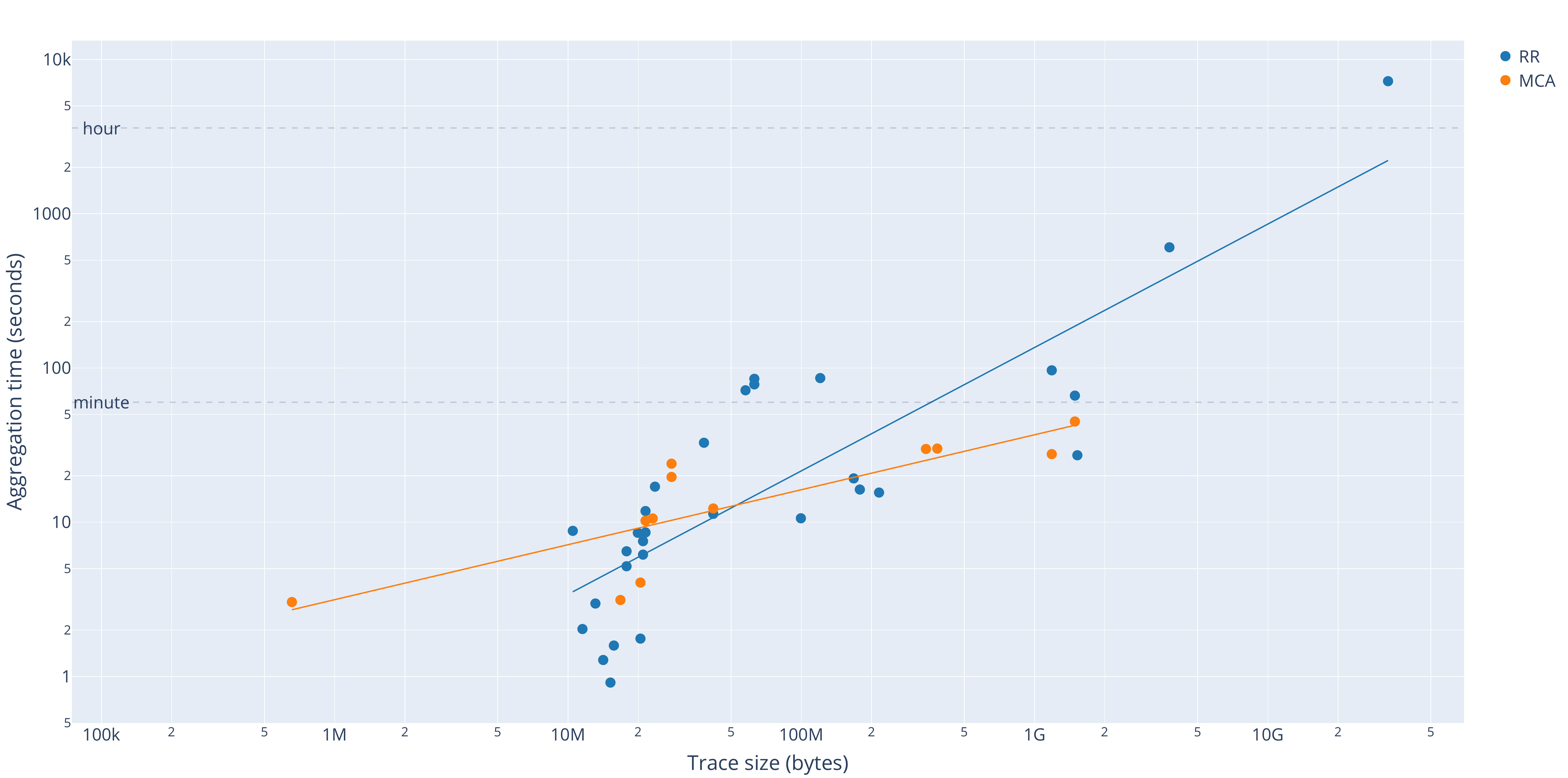}
    \caption{\pytracer aggregation performance. 
    }
    \label{fig:performance_parsing}
\end{figure}

Figure~\ref{fig:performance_parsing} shows
the postprocessing time and trace sizes produced by \pytracer. 
The linear correlation between the log of the size and the log of the time shows a significant correlation 
(p-value $\leq 10^{-8}$ for RR and p-value $\leq 10^{-3}$ for MCA).

In conclusion, while \pytracer's overhead is substantial, it remains tractable for real-life examples.

\section{Discussion}



Through our analysis of SciPy and scikit-learn, we demonstrated \pytracer's scalability, automaticity, and applicability to a wide range of Python program. The relatively low instrumentation overhead of 3.5$\times$ on average enables the analysis of real-scale applications. \pytracer's instrumentation is fully automatic and does not require manual intervention from the user. Furthermore, \pytracer works on any Python libraries, with any data format supported by pickle and any numerical noise model.

\pytracer is the first tool to enable a systematic numerical analysis of Python codes. Existing tools focus on traditional languages of the HPC community, such as C, C++, and FORTRAN. However, Python's high-level abstraction allows for quick application implementations while maintaining high-performance thanks to specialized libraries such as NumPy. Therefore, \pytracer meets the numerical analysis needs of an ever-increasing number of applications.

\pytracer allows users to pinpoint the root causes of numerical instabilities by working at the function level. Numerical stability evaluations are complex since calculations are strongly interdependent. Function-level instrumentation limits the size of the traces since only inputs and outputs of a function are dumped. Moreover, it decreases instrumentation overhead and helps the user identify instabilities quickly. Working at the function level is thus a suitable compromise between analysis conducted at the floating-point operation level and user debug prints. Indeed,  fine-grained analyses are costly and hard to interpret as they overwhelm users with information. In contrast, looking at intermediate results through print statements 
is lightweight but not reliable since developers may not systematically instrument unstable functions. Finally, the SciPy environment wraps many computation kernels written in C, such as BLAS or LAPACK. By working at the function level, \pytracer allows the user to evaluate the numerical precision of such kernels and their impact on the code without getting deep into technical implementation aspects. 

Integrating \pytracer at the early stages of the development cycle would improve the overall numerical quality of the developed code.
\pytracer's automaticity facilitates its usage and enables its integration in unit testing.
For example, \pytracer could be used in regression tests to ensure that updates do not degrade the numerical quality of a calculation. 

\pytracer can also be useful for selecting hyperparameters, as is the case in machine learning.
Indeed, it is common for data scientists to explore a range of hyperparameters and select the ones giving the best scores. With \pytracer, the user has a practical way to quantify numerical quality and understand why hyperparameters produce inconsistent results instead of tweaking hyperparameters to make solvers converge. For instance, the Face Recognition example  (Figure~\ref{fig:face_recognition_svd}) shows how numerical precision could inform the selection of the number of components in dimensionality reduction.

The visualization provides a global view of the code and interactions between computations. In addition, the Plotly dashboard helps navigate through the code easily, without the need for ad-hoc scripts to visualize data. Finally, the dashboard can be extended to support other data formats to visualize. At this stage, \pytracer currently supports n-dimensional NumPy arrays as well as native Python data. 

As we showed in Section~\ref{sec:impact_mca_modes}, using Full MCA mode leads to runtime errors that have a considerable impact on the proper functioning of the execution, which is unfortunate since these errors do not reflect actual numerical errors but are related to perturbations that should not occur. Conversely, RR mode is far more conservative since it preserves exact operations and is easier to use even though it does not simulate all perturbations. Therefore, from our experiments, we recommend the use of RR over Full MCA. More research is required to address the issues encountered with Full MCA. 

Finally, the wide range of results precision obtained from our experiments demonstrate that looking at the numerical variability is essential and that such analyses would benefit from being conducted systematically.
We showed that well-known scientific libraries are prone to numerical error even though they have been well tested for decades.
The omnipresence of these libraries in scientific codes should thus impel users to integrate numerical analysis in their development workflow in addition to testing their implementation.

\section{Related work}

Several tools have been developed to assess numerical quality, and can be divided into static and dynamic approaches. Static approaches typically analyze source code, while dynamic approaches trace floating-point computations throughout executions. A detailed review of these approaches is presented in~\cite{cherubin2020tools}. 
Numerous tools exists to trace floating-point computations for C, C++, or FORTRAN programs due to the prevalence of these languages in High-Performance Computing (HPC). 
The main tracing techniques are \textit{source-to-source}, \textit{compiler-based transformations}, and \textit{dynamic binary instrumentation}.

The source-to-source technique requires a rewriting of the application to modify floating-point types and provides a fine-grained control on the analyzed code.
CADNA~\cite{jezequel2008cadna} is a library for C, C++, and Fortran implementing  
CESTAC~\cite{vignes1993stochastic} stochastic arithmetic. Shaman~\cite{demeure_phd} is a C++ library that uses a first-order error model to propagate numerical error. 
MCAlib~\cite{frechtling2015mcalib} is a C library that implements the Monte Carlo Arithmetic (MCA)~\cite{parker1997monte} using the MPFR~\cite{fousse2007mpfr} library.
Finally, the work in~\cite{tang2016software} proposed a source-to-source framework for executing targeted code in infinite and fixed precision with and without stochastic arithmetic. The main drawback of the source-to-source approach is its lacks of scalability since rewriting large codes can be a tedious task.
 

The compiler-based approach instruments floating-point expressions at compile time
and so allows the user to automatically instrument large codebases.
Verificarlo~\cite{verificarlo} is a compiler that supports different types of arithmetic instrumentations, including MCA. 
The work in~\cite{bao2013fly} modified the GNU Compiler Collection (GCC) to track local floating-point errors across executions. pLiner~\cite{guo2020pliner} is a root-cause analyzer based on the Clang compiler that detects floating-point operations responsible for result variability using a source-to-source transformation at the Abstract Syntax Tree (AST) level to rewrite parts of code with higher precision. 
PFPSanitizer~\cite{chowdhary2020debugging,chowdhary2021parallel} is a compiler that uses parallel shadow execution to detect numerical issues using higher precision.
FLiT~\cite{sawaya2017flit} is a framework to detect variability induced by compilers and their optimizations.
The work in~\cite{wang2012development} proposes a numerical debugger based on GDB~\cite{stallman1988debugging} for Discrete Stochastic Arithmetic (DSA) on FPGA as an Eclipse plugging. Similarly, Cadtrace~\cite{jezequel2008cadna} and Shaman propose a GDB-based tool to use the CADNA and Shaman libraries with GDB, respectively.
The main limitation of the compiler-based approach is that one must have access to the source code.

Dynamic Binary Instrumentation (DBI) operates directly on executables, without the need for recompilation or manual changes. Therefore, it is applicable to any programming language.
CraftHPC~\cite{lam2013dynamic} uses DBI to detect catastrophic cancellations at runtime.
Verrou~\cite{fevotte2016verrou} is a Valgrind~\cite{nethercote2007valgrind} based tool that dynamically replaces
floating-point operations with their stochastic arithmetic counterparts. FPDebug~\cite{benz2012dynamic} uses DBI to detect numerical inaccuracies by using shadow computations with higher precision.
Herbgring~\cite{sanchez2017finding} is a Valgrind-based tool to detect
numerical instabilities. It uses a shadow memory to detect precision losses by comparison with results obtained at higher precision. It is combined with symbolic computation to backtrack the root of the error.
We can note that all methods based on a high-precision oracle suppose that computing with a large number of bits is sufficient to obtain an accurate result, which is not always true (see for example, the famous Muller's sequence~\cite{bajard1996introduction}). 
Although versatile, the DBI looses high-level information useful to understand the source-code logic.

Being agnostic to the programming language is a serious advantage of the DBI method compared to source-to-source and compiler-bases methods. However, working at the binary level makes it difficult to access high-level information needed to debug and understand the source-code logic. Conversely, the source-to-source approach provides a fine-grained control on the analyzed code, but it lacks scalability, as rewriting large codes can be a tedious task. Finally, the compiler-based approach takes advantage of the best of both by being automatic and having access to high-level information. However, like source-to-source, the compiler-based method is not suitable for analyzing closed-source libraries.

To the best of our knowledge, \pytracer is the first tool dedicated to the numerical analysis of Python code. Existing tools for tracing Python code focus on performance profiling for time (cProfile) 
or memory consumption (memprofile). 
Anteater~\cite{faust2019anteater} is a Python tracer tool to debug Python applications. 
It performs source transformations at the AST level but only deals with native numeric Python types.
Moreover, the user needs to manually tag each variable to trace. Finally, according to the authors, Anteater does not scale to large traces.
cProfile, memprofile, and Anteater use Python decorators as the primary instrumentation technique, a Python mechanism to instrument a function by adding a line over a function declaration.
While this method is appropriate when targeting specific functions, it is not feasible for large codebases where potentially unstable code sections are unknown.

\section{Conclusion}


Evaluating the numerical stability of scientific Python programs is a tedious task that requires substantial expertise. Automatic tools to help users understand the behavior of their code are therefore crucial.
In this perspective, \pytracer is the first tool to offer a complete framework for analyzing and visualizing the numerical stability of Python codes. It works by automatically instrumenting Python libraries to produce numerical traces visualized through an interactive Plotly Dash server. \pytracer does not require manual code modification, and we tested it with major scientific libraries such as NumPy, SciPy, and Scikit-learn. Our results show that \pytracer is scalable, with an average instrumentation slowdown of $3.5\times$. Moreover, \pytracer's visualization interface remains responsive even with traces of up to 1TB. 
In our experiments, \pytracer enabled the systematic evaluation of 40 examples from the SciPy and scikit-learn libraries. 

In future work, we plan to extend the post-processing analysis to aggregate traces when the execution flow path diverges by leveraging techniques used in code coverage analysis. Moreover, we would like to extend \pytracer to instrument elementary arithmetic operations which can be helpful to more profound analyses. Finally, although \pytracer supports multiple noise models, we only tested it with Monte Carlo Arithmetic. Therefore, we would like to compare different models against MCA. 
\pytracer is available at \href{https://github.com/big-data-lab-team/pytracer}{github://big-data-lab-team/pytracer} under MIT License.

\bibliographystyle{abbrv}
\bibliography{paper}

\appendix

\end{document}

%% file: href.tex
\newcommand{\discreteCosineRF}{ \href{https://docs.scipy.org/doc/scipy/tutorial/fft.html\#discrete-cosine-transforms}{discrete\_cosine\_transform} }
\newcommand{\fftOneDRf}{
\href{https://docs.scipy.org/doc/scipy/tutorial/fft.html\#d-discrete-fourier-transforms}{fft\_1D}
}
    
\newcommand{\fftTwoDRf}{
\href{https://docs.scipy.org/doc/scipy/tutorial/fft.html\#and-n-d-discrete-fourier-transforms}{fft\_2D}
}

\newcommand{\interOneDRf}{
\href{https://docs.scipy.org/doc/scipy/tutorial/interpolate.html\#d-interpolation-interp1d}{interpolation\_1D} 
}

\newcommand{\multiRf}{
\href{https://docs.scipy.org/doc/scipy/tutorial/interpolate.html\#multivariate-data-interpolation-griddata}{multivariate\_data} 
}

\newcommand{\bsplineRf}{
\href{https://docs.scipy.org/doc/scipy/tutorial/signal.html\#b-splines}{bspline}  
}

\newcommand{\splineOneDRf}{
    \href{https://docs.scipy.org/doc/scipy/tutorial/interpolate.html\#spline-interpolation-in-1-d-procedural-interpolate-splxxx}{
    spline\_1D} 
}

\newcommand{\splineTwoDRf}{
    \href{https://docs.scipy.org/doc/scipy/tutorial/interpolate.html\#d-spline-representation-procedural-bisplrep}{
    spline\_2D} 
}

\newcommand{\bfgsRf}{
    \href{https://docs.scipy.org/doc/scipy/tutorial/optimize.html\#broyden-fletcher-goldfarb-shanno-algorithm-method-bfgs}{broyden\_fletcher\_goldfarb\_shanno} 
}

\newcommand{\globalRf}{
    \href{https://docs.scipy.org/doc/scipy/tutorial/optimize.html\#global-optimization}{
    global\_optimization} 
}

\newcommand{\slsqpRf}{
    \href{https://docs.scipy.org/doc/scipy/tutorial/optimize.html\#sequential-least-squares-programming-slsqp-algorithm-method-slsqp}{
    slsqp} 
}
    
\newcommand{\lsmRf}{
    \href{https://docs.scipy.org/doc/scipy/tutorial/optimize.html\#least-squares-minimization-least-squares}{
    least\_square\_minimization} 
}

\newcommand{\nelderRf}{
    \href{https://docs.scipy.org/doc/scipy/tutorial/optimize.html\#nelder-mead-simplex-algorithm-method-nelder-mead}{nelder\_mead\_simplex}
}

\newcommand{\ncgRf}{
    \href{https://docs.scipy.org/doc/scipy/tutorial/optimize.html\#newton-conjugate-gradient-algorithm-method-newton-cg}{
    newton\_conjugate\_gradient}  
}

\newcommand{\rootRf}{
    \href{https://docs.scipy.org/doc/scipy/tutorial/optimize.html\#sets-of-equations}{
    root\_findings}  
}

\newcommand{\rootLargRf}{
    \href{https://docs.scipy.org/doc/scipy/tutorial/optimize.html\#root-finding-for-large-problems}{
    root\_finding\_large} 
}

\newcommand{\rootLargePredRf}{
    \href{https://docs.scipy.org/doc/scipy/tutorial/optimize.html\#still-too-slow-preconditioning}{
    root\_finding\_large\_preconditioned} 
}

\newcommand{\trustExactRf}{
    \href{https://docs.scipy.org/doc/scipy/tutorial/optimize.html\#trust-region-nearly-exact-algorithm-method-trust-exact}{
    trust\_region\_exact} 
}

\newcommand{\trustTruncRf}{
    \href{https://docs.scipy.org/doc/scipy/tutorial/optimize.html\#trust-region-truncated-generalized-lanczos-conjugate-gradient-algorithm-method-trust-krylov}{
    trust\_region\_truncated\_generalized\_lanczos} 
}

\newcommand{\trustNCGRf}{
    \href{https://docs.scipy.org/doc/scipy/tutorial/optimize.html\#trust-region-newton-conjugate-gradient-algorithm-method-trust-ncg}{
    trust\_region\_newton\_conjugate\_gradient} 
}

\newcommand{\AdaboostRf}{
    \href{https://scikit-learn.org/stable/auto_examples/ensemble/plot_adaboost_regression.html}{AdaBoost} 
}

\newcommand{\brrRf}{
    \href{https://scikit-learn.org/stable/auto_examples/linear_model/plot_bayesian_ridge.html}{Bayesian Ridge Regression} 
}

\newcommand{\onlineClassifierComparisonRf}{
    \href{https://scikit-learn.org/stable/auto_examples/linear_model/plot_sgd_comparison.html}{Online classifier comparison} 
}

\newcommand{\kmeansRf}{
    \href{https://scikit-learn.org/stable/auto_examples/cluster/plot_cluster_iris.html}{K-Means Clustering}  
}

\newcommand{\covarianceRf}{
    \href{https://scikit-learn.org/stable/auto_examples/covariance/plot_mahalanobis_distances.html}{Covariance estimation}  
}

\newcommand{\decisionRf}{
    \href{https://scikit-learn.org/stable/auto_examples/tree/plot_tree_regression.html}{Decision Tree Regression}
}

\newcommand{\digitsRf}{
    \href{https://scikit-learn.org/stable/auto_examples/linear_model/plot_sgd_comparison.html}{Digits Classification} 
}

\newcommand{\faceRf}{
    \href{https://scikit-learn.org/stable/auto_examples/decomposition/plot_faces_decomposition.html}{Face Recognition} 
}

\newcommand{\penaltyRf}{
    \href{https://scikit-learn.org/stable/auto_examples/linear_model/plot_logistic_l1_l2_sparsity.html}{L1 Penalty} 
}

\newcommand{\lassoRf}{
    \href{https://scikit-learn.org/stable/auto_examples/linear_model/plot_lasso_and_elasticnet.html}{Lasso and Elastic Net} 
}

\newcommand{\hyperplaneRf}{
    \href{https://scikit-learn.org/stable/auto_examples/linear_model/plot_sgd_separating_hyperplane.html}{Separating hyperplane} 
}

\newcommand{\mnistRf}{
    \href{https://scikit-learn.org/stable/auto_examples/linear_model/plot_sparse_logistic_regression_mnist.html}{MNIST classification} 
}

\newcommand{\multitaskRf}{
    \href{https://scikit-learn.org/stable/auto_examples/linear_model/plot_multi_task_lasso_support.html}{Multitask Lasso}  
}

\newcommand{\ompRf}{
    \href{https://scikit-learn.org/stable/auto_examples/linear_model/plot_omp.html}{Orthogonal Matching} 
}

\newcommand{\pcaRf}{
    \href{https://scikit-learn.org/stable/auto_examples/decomposition/plot_pca_iris.html}{PCA decomposition}  
}

\newcommand{\robustRf}{
    \href{https://scikit-learn.org/stable/auto_examples/linear_model/plot_ransac.html}{Robust Linear Regression} 
}

\newcommand{\toyRf}{
    \href{https://scikit-learn.org/stable/auto_examples/cluster/plot_segmentation_toy.html\#sphx-glr-auto-examples-cluster-plot-segmentation-toy-py}{Segmentation toy}  
}

\newcommand{\svmRf}{
    \href{https://scikit-learn.org/stable/auto_examples/svm/plot_separating_hyperplane_unbalanced.html\#sphx-glr-auto-examples-svm-plot-separating-hyperplane-unbalanced-py}{SVM}  
}

\newcommand{\tomographyRf}{
    \href{https://scikit-learn.org/stable/auto_examples/applications/plot_tomography_l1_reconstruction.html\#sphx-glr-auto-examples-applications-plot-tomography-l1-reconstruction-py}{
    Tomography} 
}

\newcommand{\weightedRf}{
    \href{https://scikit-learn.org/stable/auto_examples/linear_model/plot_sgd_weighted_samples.html\#sphx-glr-auto-examples-linear-model-plot-sgd-weighted-samples-py}{SGD weighted samples} 
}

%% file: paper.bbl
\begin{thebibliography}{10}

\bibitem{bajard1996introduction}
J.-C. Bajard, D.~Michelucci, J.-M. Moreau, and J.-M. Muller.
\newblock {Introduction to the Special Issue "Real Numbers and Computers"}.
\newblock In {\em The Journal of Universal Computer Science}, pages 436--438.
  Springer, 1996.

\bibitem{bao2013fly}
T.~Bao and X.~Zhang.
\newblock On-the-fly detection of instability problems in floating-point
  program execution.
\newblock In {\em Proceedings of the 2013 ACM SIGPLAN international conference
  on Object oriented programming systems languages \& applications}, pages
  817--832, 2013.

\bibitem{benz2012dynamic}
F.~Benz, A.~Hildebrandt, and S.~Hack.
\newblock A dynamic program analysis to find floating-point accuracy problems.
\newblock {\em ACM SIGPLAN Notices}, 47(6):453--462, 2012.

\bibitem{BFGS}
C.~G. BROYDEN.
\newblock {The Convergence of a Class of Double-rank Minimization Algorithms:
  2. The New Algorithm}.
\newblock {\em IMA Journal of Applied Mathematics}, 6(3):222--231, 09 1970.

\bibitem{chatelain2019outils}
Y.~Chatelain.
\newblock {\em Outils de d{\'e}bogage et d'optimisation des calculs flottants
  dans le contexte HPC}.
\newblock PhD thesis, Universit{\'e} Paris-Saclay, 2019.

\bibitem{chatelain2019automatic}
Y.~Chatelain, E.~Petit, P.~de~Oliveira~Castro, G.~Lartigue, and D.~Defour.
\newblock Automatic exploration of reduced floating-point representations in
  iterative methods.
\newblock In {\em European Conference on Parallel Processing}, pages 481--494.
  Springer, 2019.

\bibitem{cherubin2020tools}
S.~Cherubin and G.~Agosta.
\newblock Tools for reduced precision computation: A survey.
\newblock {\em ACM Computing Surveys (CSUR)}, 53(2):1--35, 2020.

\bibitem{chowdhary2020debugging}
S.~Chowdhary, J.~P. Lim, and S.~Nagarakatte.
\newblock Debugging and detecting numerical errors in computation with posits.
\newblock In {\em Proceedings of the 41st ACM SIGPLAN Conference on Programming
  Language Design and Implementation}, pages 731--746, 2020.

\bibitem{chowdhary2021parallel}
S.~Chowdhary and S.~Nagarakatte.
\newblock Parallel shadow execution to accelerate the debugging of numerical
  errors.
\newblock In {\em 2021 The ACM Joint European Software Engineering Conference
  and Symposium on the Foundations of Software Engineering (ESEC/FSE)}, 2021.

\bibitem{demeure_phd}
N.~Demeure.
\newblock {\em Compromise between precision and performance in high performance
  computing.}
\newblock Theses, {{\'E}cole Normale sup{\'e}rieure Paris-Saclay}, Jan. 2021.

\bibitem{verificarlo}
C.~Denis, P.~de~Oliveira~Castro, and E.~Petit.
\newblock Verificarlo: Checking floating point accuracy through monte carlo
  arithmetic.
\newblock In {\em 23nd {IEEE} Symposium on Computer Arithmetic, {ARITH} 2016,
  Silicon Valley, CA, USA, July 10-13, 2016}, pages 55--62, 2016.

\bibitem{faust2019anteater}
R.~Faust, K.~Isaacs, W.~Z. Bernstein, M.~Sharp, and C.~Scheidegger.
\newblock Anteater: Interactive visualization for program understanding.
\newblock {\em arXiv preprint arXiv:1907.02872}, 2019.

\bibitem{fevotte2016verrou}
F.~F{\'e}votte and B.~Lathuiliere.
\newblock Verrou: a cestac evaluation without recompilation.
\newblock {\em SCAN 2016}, page~47, 2016.

\bibitem{fousse2007mpfr}
L.~Fousse, G.~Hanrot, V.~Lef{\`e}vre, P.~P{\'e}lissier, and P.~Zimmermann.
\newblock Mpfr: A multiple-precision binary floating-point library with correct
  rounding.
\newblock {\em ACM Transactions on Mathematical Software (TOMS)}, 33(2):13--es,
  2007.

\bibitem{frechtling2015mcalib}
M.~Frechtling and P.~H. Leong.
\newblock Mcalib: Measuring sensitivity to rounding error with monte carlo
  programming.
\newblock {\em ACM Transactions on Programming Languages and Systems (TOPLAS)},
  37(2):1--25, 2015.

\bibitem{gould1999solving}
N.~I. Gould, S.~Lucidi, M.~Roma, and P.~L. Toint.
\newblock Solving the trust-region subproblem using the lanczos method.
\newblock {\em SIAM Journal on Optimization}, 9(2):504--525, 1999.

\bibitem{guo2020pliner}
H.~Guo, I.~Laguna, and C.~Rubio-Gonz{\'a}lez.
\newblock pliner: isolating lines of floating-point code for compiler-induced
  variability.
\newblock In {\em SC20: International Conference for High Performance
  Computing, Networking, Storage and Analysis}, pages 1--14. IEEE, 2020.

\bibitem{halko2011finding}
N.~Halko, P.-G. Martinsson, and J.~A. Tropp.
\newblock Finding structure with randomness: Probabilistic algorithms for
  constructing approximate matrix decompositions.
\newblock {\em SIAM review}, 53(2):217--288, 2011.

\bibitem{harris2020array}
C.~R. Harris, K.~J. Millman, S.~J. van~der Walt, R.~Gommers, P.~Virtanen,
  D.~Cournapeau, E.~Wieser, J.~Taylor, S.~Berg, N.~J. Smith, et~al.
\newblock Array programming with numpy.
\newblock {\em Nature}, 585(7825):357--362, 2020.

\bibitem{plotly}
P.~T. Inc.
\newblock Collaborative data science, 2015.

\bibitem{jezequel2008cadna}
F.~J{\'e}z{\'e}quel and J.-M. Chesneaux.
\newblock Cadna: a library for estimating round-off error propagation.
\newblock {\em Computer Physics Communications}, 178(12):933--955, 2008.

\bibitem{kiar2020comparing}
G.~Kiar, P.~de~Oliveira~Castro, P.~Rioux, E.~Petit, S.~T. Brown, A.~C. Evans,
  and T.~Glatard.
\newblock Comparing perturbation models for evaluating stability of
  neuroimaging pipelines.
\newblock {\em The International Journal of High Performance Computing
  Applications}, page 1094342020926237, 2020.

\bibitem{kowalik1968analysis}
J.~Kowalik and J.~Morrison.
\newblock Analysis of kinetic data for allosteric enzyme reactions as a
  nonlinear regression problem.
\newblock {\em Mathematical Biosciences}, 2(1-2):57--66, 1968.

\bibitem{kraft1988software}
D.~Kraft et~al.
\newblock A software package for sequential quadratic programming.
\newblock Technical report, Deutsche Forschungs- und Versuchsanstalt fuer Luft-
  und Raumfahrt e.V., 1988.

\bibitem{lam2013dynamic}
M.~O. Lam, J.~K. Hollingsworth, and G.~Stewart.
\newblock Dynamic floating-point cancellation detection.
\newblock {\em Parallel Computing}, 39(3):146--155, 2013.

\bibitem{lattner2008llvm}
C.~Lattner.
\newblock Llvm and clang: Next generation compiler technology.
\newblock In {\em The BSD conference}, volume~5, 2008.

\bibitem{li1993centering}
Y.~Li.
\newblock Centering, trust region, reflective techniques for nonlinear
  minimization subject to bounds.
\newblock Technical report, Cornell University, 1993.

\bibitem{nakatsukasa2013stable}
Y.~Nakatsukasa and N.~J. Higham.
\newblock Stable and efficient spectral divide and conquer algorithms for the
  symmetric eigenvalue decomposition and the svd.
\newblock {\em SIAM Journal on Scientific Computing}, 35(3):A1325--A1349, 2013.

\bibitem{nethercote2007valgrind}
N.~Nethercote and J.~Seward.
\newblock Valgrind: a framework for heavyweight dynamic binary instrumentation.
\newblock {\em ACM Sigplan notices}, 42(6):89--100, 2007.

\bibitem{nocedal2006numerical}
J.~Nocedal and S.~Wright.
\newblock {\em Numerical optimization}.
\newblock Springer Science \& Business Media, 2006.

\bibitem{parker1997monte}
D.~S. Parker.
\newblock {\em Monte Carlo arithmetic: exploiting randomness in floating-point
  arithmetic}.
\newblock University of California (Los Angeles). Computer Science Department,
  1997.

\bibitem{paszke2019pytorch}
A.~Paszke, S.~Gross, F.~Massa, A.~Lerer, J.~Bradbury, G.~Chanan, T.~Killeen,
  Z.~Lin, N.~Gimelshein, L.~Antiga, et~al.
\newblock Pytorch: An imperative style, high-performance deep learning library.
\newblock {\em Advances in neural information processing systems},
  32:8026--8037, 2019.

\bibitem{pedregosa2011scikit}
F.~Pedregosa, G.~Varoquaux, A.~Gramfort, V.~Michel, B.~Thirion, O.~Grisel,
  M.~Blondel, P.~Prettenhofer, R.~Weiss, V.~Dubourg, et~al.
\newblock Scikit-learn: Machine learning in python.
\newblock {\em the Journal of machine Learning research}, 12:2825--2830, 2011.

\bibitem{Platt99probabilisticoutputs}
J.~C. Platt.
\newblock Probabilistic outputs for support vector machines and comparisons to
  regularized likelihood methods.
\newblock In {\em ADVANCES IN LARGE MARGIN CLASSIFIERS}, pages 61--74. MIT
  Press, 1999.

\bibitem{sanchez2017finding}
A.~Sanchez-Stern, P.~Panchekha, S.~Lerner, and Z.~Tatlock.
\newblock Finding root causes of floating point error with herbgrind.
\newblock {\em arXiv preprint arXiv:1705.10416}, 2017.

\bibitem{sawaya2017flit}
G.~Sawaya, M.~Bentley, I.~Briggs, G.~Gopalakrishnan, and D.~H. Ahn.
\newblock Flit: Cross-platform floating-point result-consistency tester and
  workload.
\newblock In {\em 2017 IEEE international symposium on workload
  characterization (IISWC)}, pages 229--238. IEEE, 2017.

\bibitem{singer2009nelder}
S.~Singer and J.~Nelder.
\newblock Nelder-mead algorithm.
\newblock {\em Scholarpedia}, 4(7):2928, 2009.

\bibitem{sohier2018confidence}
D.~Sohier, P.~D.~O. Castro, F.~F{\'e}votte, B.~Lathuili{\`e}re, E.~Petit, and
  O.~Jamond.
\newblock Confidence intervals for stochastic arithmetic.
\newblock {\em arXiv preprint arXiv:1807.09655}, 2018.

\bibitem{stallman1988debugging}
R.~Stallman, R.~Pesch, S.~Shebs, et~al.
\newblock Debugging with gdb.
\newblock {\em Free Software Foundation}, 675, 1988.

\bibitem{tang2016software}
E.~Tang, X.~Zhang, N.~T. M{\"u}ller, Z.~Chen, and X.~Li.
\newblock Software numerical instability detection and diagnosis by combining
  stochastic and infinite-precision testing.
\newblock {\em IEEE Transactions on Software Engineering}, 43(10):975--994,
  2016.

\bibitem{vignes1993stochastic}
J.~Vignes.
\newblock A stochastic arithmetic for reliable scientific computation.
\newblock {\em Mathematics and computers in simulation}, 35(3):233--261, 1993.

\bibitem{virtanen2020scipy}
P.~Virtanen, R.~Gommers, T.~E. Oliphant, M.~Haberland, T.~Reddy, D.~Cournapeau,
  E.~Burovski, P.~Peterson, W.~Weckesser, J.~Bright, et~al.
\newblock Scipy 1.0: fundamental algorithms for scientific computing in python.
\newblock {\em Nature methods}, 17(3):261--272, 2020.

\bibitem{wang2012development}
K.~Wang.
\newblock Development of a graphical numerical accuracy debugger based on an
  fpga computing system.
\newblock Master's thesis, Universität Stuttgart, 2012.

\end{thebibliography}
